\documentclass[longauth]{aa}  
\usepackage{graphicx}
\usepackage[varg]{txfonts}
\bibpunct{(}{)}{;}{a}{ }{,}

\usepackage[switch]{lineno}

\begin{document}

\title{Extended VHE $\gamma$-ray emission towards SGR1806$-$20, LBV\,1806$-$20, and stellar cluster Cl*\,1806$-$20}
\titlerunning{VHE $\gamma$-ray emission towards SGR\,1806$-$20, LBV\,1806$-$20 and Cl*\,1806$-$20}


\author{\small H.E.S.S. Collaboration
\and H.~Abdalla \inst{1}
\and A.~Abramowski \inst{2}
\and F.~Aharonian \inst{3,4,5}
\and F.~Ait Benkhali \inst{3}
\and A.G.~Akhperjanian \inst{6,5}
\and E.O.~Ang\"uner \inst{7}
\and M.~Arrieta \inst{15}
\and P.~Aubert \inst{24}
\and M.~Backes \inst{8}
\and A.~Balzer \inst{9}
% ...
}

\institute{
Centre for Space Research, North-West University, Potchefstroom 2520, South Africa \and 
Universit\"at Hamburg, Institut f\"ur Experimentalphysik, Luruper Chaussee 149, D 22761 Hamburg, Germany \and 
Max-Planck-Institut f\"ur Kernphysik, P.O. Box 103980, D 69029 Heidelberg, Germany \and 
Dublin Institute for Advanced Studies, 31 Fitzwilliam Place, Dublin 2, Ireland \and 
National Academy of Sciences of the Republic of Armenia,  Marshall Baghramian Avenue, 24, 0019 Yerevan, Republic of Armenia  \and
Yerevan Physics Institute, 2 Alikhanian Brothers St., 375036 Yerevan, Armenia \and
Institut f\"ur Physik, Humboldt-Universit\"at zu Berlin, Newtonstr. 15, D 12489 Berlin, Germany \and
University of Namibia, Department of Physics, Private Bag 13301, Windhoek, Namibia \and
GRAPPA, Anton Pannekoek Institute for Astronomy, University of Amsterdam,  Science Park 904, 1098 XH Amsterdam, The Netherlands \and
Department of Physics and Electrical Engineering, Linnaeus University,  351 95 V\"axj\"o, Sweden \and
Institut f\"ur Theoretische Physik, Lehrstuhl IV: Weltraum und Astrophysik, Ruhr-Universit\"at Bochum, D 44780 Bochum, Germany \and
GRAPPA, Anton Pannekoek Institute for Astronomy and Institute of High-Energy Physics, University of Amsterdam,  Science Park 904, 1098 XH Amsterdam, The Netherlands \and
Institut f\"ur Astro- und Teilchenphysik, Leopold-Franzens-Universit\"at Innsbruck, A-6020 Innsbruck, Austria \and
School of Chemistry \& Physics, University of Adelaide, Adelaide 5005, Australia \and
LUTH, Observatoire de Paris, PSL Research University, CNRS, Universit\'e Paris Diderot, 5 Place Jules Janssen, 92190 Meudon, France \and
Sorbonne Universit\'es, UPMC Universit\'e Paris 06, Universit\'e Paris Diderot, Sorbonne Paris Cit\'e, CNRS, Laboratoire de Physique Nucl\'eaire et de Hautes Energies (LPNHE), 4 place Jussieu, F-75252, Paris Cedex 5, France \and
Laboratoire Univers et Particules de Montpellier, Universit\'e Montpellier, CNRS/IN2P3,  CC 72, Place Eug\`ene Bataillon, F-34095 Montpellier Cedex 5, France \and
DSM/Irfu, CEA Saclay, F-91191 Gif-Sur-Yvette Cedex, France \and
Astronomical Observatory, The University of Warsaw, Al. Ujazdowskie 4, 00-478 Warsaw, Poland \and
Aix Marseille Universit\'e, CNRS/IN2P3, CPPM UMR 7346,  13288 Marseille, France \and
Instytut Fizyki J\c{a}drowej PAN, ul. Radzikowskiego 152, 31-342 Krak{\'o}w, Poland \and
Funded by EU FP7 Marie Curie, grant agreement No. PIEF-GA-2012-332350,  \and
School of Physics, University of the Witwatersrand, 1 Jan Smuts Avenue, Braamfontein, Johannesburg, 2050 South Africa \and
Laboratoire d'Annecy-le-Vieux de Physique des Particules, Universit\'{e} Savoie Mont-Blanc, CNRS/IN2P3, F-74941 Annecy-le-Vieux, France
% ...
}

\offprints{H.E.S.S.~Collaboration,\\
\email{contact.hess@hess-experiment.eu};\\
$\star$ Corresponding authors: G.\,Rowell, M.\,de\,Naurois}

\date{Received: September 12 April 2016 / Accepted: 11 May 2016}

\abstract {
  Using the High Energy Spectroscopic System (H.E.S.S.) telescopes we have discovered a steady and extended very high-energy (VHE) 
  $\gamma$-ray source 
  towards the luminous blue variable candidate LBV\,1806$-$20, massive stellar cluster Cl*\,1806$-$20, and magnetar SGR\,1806$-$20. 
  The new VHE source, HESS\,J1808$-$204, was detected at a statistical significance of $>6\sigma$ (post-trial) with a photon 
  flux normalisation
  $(2.9 \pm 0.4_{\rm stat} \pm 0.5_{\rm sys})\times 10^{-13}$\,ph\,cm$^{-2}$\,s$^{-1}$\,TeV$^{-1}$ at 1\,TeV and a power-law photon
  index of $2.3\pm0.2_{\rm stat}\pm 0.3_{\rm sys}$. The luminosity of this source (0.2 to 10 TeV; scaled to distance $d$=8.7\,kpc) 
  is $L_{\rm VHE}\sim1.6 \times 10^{34}(d/{\rm 8.7\, kpc})^2$\,erg\,s$^{-1}$. The VHE $\gamma$-ray emission is 
  extended and is well fit by a single Gaussian with statistical standard deviation of $0.095^\circ \pm 0.015^\circ$. 
  This extension is similar to that of the synchrotron radio nebula 
  G10.0$-$0.3, which is thought to be powered by 
  LBV\,1806$-$20. The VHE $\gamma$-ray luminosity could be provided by the stellar wind luminosity of LBV\,1806$-$20 by itself 
  and/or the massive star members of Cl*\,1806$-$20. Alternatively, magnetic dissipation (e.g. via reconnection) 
  from SGR\,1806$-$20 can 
  potentially account for the VHE luminosity.
  The origin and hadronic and/or leptonic nature of the accelerated particles responsible for HESS\,J1808$-$204 is not 
  yet clear. If associated with SGR\,1806$-$20, the potentially young age of the magnetar (650\,yr) can be used to infer 
  the transport limits of these particles to match the VHE source size. 
  This discovery provides new
  interest in the potential for high-energy particle acceleration from magnetars, massive stars, and/or stellar clusters.}

\maketitle 

\section{Introduction}
 The magnetar \object{SGR\,1806$-$20} (e.g. \citealt{laros:1986}) is one of the most prominent and burst-active 
 soft gamma repeaters (SGRs). It is best known for its giant flare of 27 December 2004 \citep{hurley:2005}, one of the 
 strongest $\gamma$-ray outbursts recorded with a luminosity reaching $L\sim10^{47}$\,erg\,s$^{-1}$ (from radio to hard X-ray energies).
 SGR\,1806$-$20 is also a member of the massive stellar cluster \object{Cl*\,1806$-$20} \citep{fuchs:1999,eikenberry:2001,figer:2005}. 
 This cluster  harbours (within a $0.5^\prime$ radius) a number of energetic stars such as four Wolf-Rayet (WR) stars, 
 five O-type stars, and a rare luminous blue variable candidate (cLBV), \object{LBV\,1806$-$20} \citep{kulkarni:1995,vankerwijk:1995}. 

 In addition to the giant flare, a number of less intense 
 flares ($L\sim10^{40}-10^{43}$\,erg\,s$^{-1}$) from SGR\,1806$-$20 have also been seen in the past decade. 
 The energy source for the flares is often attributed to magnetic torsion acting on and leading to deformation 
 of the neutron star (NS) 
 surface \citep{duncan:1992,paczynski:1992}. 
 The giant flare luminosity may instead result from accretion events onto a quark star 
   (see e.g. \citealt{ouyed:2007}).

 Magnetars are NSs with intense magnetic fields of order $B\sim10^{14}-10^{15}$\,G. They represent one of nature's extreme 
 astrophysical objects. Compared to canonical NSs (with $B\sim10^{10}$ to $10^{13}$\,G), magnetars exhibit slower 
 spin rates ($P$ of the order of a few seconds) 
 but considerably faster spin-down rates ($\dot{P}\sim10^{-11}-10^{-9}$\,s\,s$^{-1}$; see magnetar catalogue by \citealt{olausen:2014}).
 For SGR\,1806$-$20, the values $P=7.6$\,s and $\dot{P}=7.5\times 10^{-10}$\,s\,s$^{-1}$ have been determined \citep{nakagawa:2009b}, 
 suggesting a spin-down power of $L_{\rm SD}\sim10^{34}-10^{35}$\,erg\,s$^{-1}$ 
 (see also \citealt{mereghetti:2011} and \citealt{younes:2015}). 
 This appears insufficient to account for the quiescent 
 (unpulsed) X-ray emission with luminosity of $L_{\rm X}\sim10^{35}$\,erg\,s$^{-1}$. 

 The non-flaring X-ray emission is likely related to the decay of the intense magnetic field, 
 which can theoretically yield a luminosity of $L_{\rm B}\sim10^{35}-10^{36}$\,erg\,s$^{-1}$ \citep{zhang:2003}. This X-ray emission
 to which a variety of thermal and/or non-thermal models were fit 
 is in fact variable and increased by a factor of 2 to 3 
 around the giant flare epoch of 2004/2005 \citep{mereghetti:2007,goetz:2007,esposito:2007,nakagawa:2009a}. 
 The quiescent X-ray emission is point-like as viewed by {\em Chandra} ($\lesssim 3^{\prime \prime}$), and a faint extension 
 out to $\sim1^\prime$ due to scattering by dust \citep{kaplan:2002,svirski:2011,vigano:2014} has been noticed in the two years following
 the giant flare of late 2004. 

 Interpretation of the $E<10$\,keV X-ray emission so far has centred on hot thermal gas with an additional non-thermal 
 component arising 
 from inverse-Compton (IC) scattering of NS thermal 
 photons by NS wind electron/positron pairs. For $E>10$\,keV, the possibility of super-heated thermal Bremsstrahlung  
 ($kT\sim100$\,keV), synchrotron, and IC emission has been debated 
 (see reviews by \citealt{harding:2006} and \citealt{mereghetti:2011}).
 
 LBV\,1806$-$20 may be one of the most luminous ($L>5\times10^6$L$_\odot$) and massive ($M\sim100$\,M$_\odot$) stars known 
 \citep{eikenberry:2004,clark:2005} although the possibility of a binary system has been suggested \citep{figer:2004}. 
 The Cl*\,1806$-$20 
 cluster age and combined stellar mass have been estimated at $3-4$\,Myr and $>$2000\,M$_\odot$ respectively. 
 However, SGR\,1806$-$20 appears to 
 be much younger with age $\sim$650\,years \citep{tendulkar:2012} based on proper motion of the magnetar and cluster member stars.
 
 Given the high-mass loss rates associated with WR stars 
 ($>10^{-5}\,$M$_{\odot}\,$yr$^{-1}$) and the even higher rates for LBVs (e.g. \citealt{clark:2005}), 
 the combined stellar wind 
 kinetic energy in Cl*\,1806$-$20 could reach $L_{\rm w}>10^{38}$\,erg\,s$^{-1}$.
 Cl*\,1806$-$20 is enveloped in a synchrotron radio nebula (G10.0$-$0.3) with luminosity of $L_{\rm Rad}\sim10^{32}$\,erg\,s$^{-1}$
 (for $d=8.7$\,kpc, see below) extending over $\sim\,9^\prime\times6^\prime$ in size \citep{kulkarni:1994}. 
 Originally linked to a supernova remnant (SNR), G10.0$-$0.3 is now believed to be powered by the intense stellar 
 wind from LBV\,1806$-$20 
 where the synchrotron flux of the nebula clearly peaks \citep{gaensler:2001,kaplan:2002}. The third source catalogue from {\em Fermi-LAT}  
 \citep{fermi3FGL} reports a confused\footnote{Potentially contaminated by Galactic diffuse and/or strong neighbour source 
   emission, especially at sub-GeV energies.} GeV $\gamma$-ray source, \object{3FGL\,J1809.2$-$2016c},
 towards the HII region G10.2$-$0.3 about $12^\prime$ to the Galactic north-west of 
 Cl*\,1806$-$20. G10.2$-$0.3 is a part of the giant HII complex W31 extending farther to the Galactic north \citep{corbel:2004}
 (see online Fig.~\ref{fig:ir_image}). 
 Distance estimates for Cl*\,1806$-$20 are in a wide range of 6 to 19\,kpc, based on a variety of
 techniques 
 \citep{corbel:2004,mcclure:2005,bibby:2008,svirski:2011}. We have adopted here the 8.7$^{+1.8}_{-1.5}$\,kpc 
 distance from \citet{bibby:2008} who used stellar spectra and inferred luminosities of the Cl*\,1806$-$20 cluster members.
 
 Very high-energy (VHE) $\gamma$-ray emission has so far not been identified or associated with magnetars \citep{magic-magnetars},
 despite the theoretical grounds for multi-TeV particle acceleration from or around them  \citep{zhang:2003b,arons:2003}
 with subsequent $\gamma$-ray and neutrino emission (e.g. \citealt{zhang:2003b,ioka:2005,liu:2010}).
 Massive stars and clusters have been suggested as multi-GeV particle accelerators (e.g. \citealt{montmerle:1979,volk:1982, 
   eichler:1993,domingo:2006,bednarek:2007,reimer:2006}) and several extended
 VHE $\gamma$-ray sources have been found towards them \citep{tevj2032,hesswd2,hesswd1}.
 The high luminosities of SGR\,1806$-$20, LBV\,1806$-$20, and Cl*\,1806$-$20, as well as the non-thermal radio, 
 and hard X-ray emission seen towards these objects
 have motivated our search for VHE $\gamma$-ray emission with H.E.S.S.
 \begin{figure}[t]
   \centering 
    \includegraphics[width=0.5\textwidth]{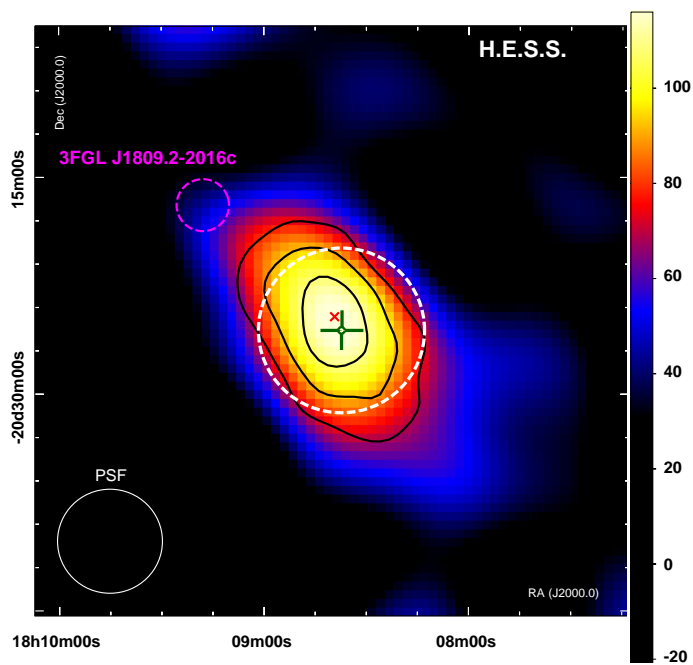}
   \caption{H.E.S.S. exposure-corrected excess counts image of HESS\,J1808$-$204 towards the stellar cluster Cl*\,1806$-$20 (red
cross), containing SGR\,1806$-$20 and LBV\,1806$-$20. The image is Gaussian smoothed with 
     standard deviation 0.06$^\circ$ corresponding to the 68\% containment radius of the H.E.S.S. analysis 
     PSF, which is also indicated by the white
     circle at the bottom left corner. 
     Pre-trial excess significance contours (6, 5, 4$\sigma$ levels) 
     for an integration radius 
     of 0.1$^\circ$ are indicated by black solid lines. The dark green solid point with error bars (1$\sigma$ statistical) and white
     dashed ellipse represent the fitted location and radius of the intrinsic Gaussian source model. The 68\% location error of 
     3FGL\,J1809.2$-$2016c ({\em Fermi-LAT} GeV source) is indicated by the magenta dashed ellipse.}
   \label{fig:tevimage}
 \end{figure}

 \section{H.E.S.S. VHE $\gamma$-ray observations, analysis, and results}

 VHE $\gamma$-ray observations have been carried out with the  High Energy Stereoscopic System (H.E.S.S.) array  
 of five imaging atmospheric Cherenkov telescopes 
 (IACTs) located in the Khomas Highland of
 Namibia ($16^\circ 30^\prime 00^{\prime \prime}$ E $23^\circ 16^\prime 18^{\prime \prime}$ S; 1800\,m above sea level). 
 The fifth telescope was added in 2012, but the data analysed here precede this installation, thus making
 use of the four original IACTs (with mirror area 107\,m$^2$).
 These IACTs provide a stereoscopic view of extensive air showers (EAS) for
 reconstruction of $\gamma$-ray primary arrival direction and energy \citep{hesscrab}. An event-by-event angular resolution of
 $0.06^\circ$ (68\% containment radius), energy resolution $\Delta E/E \lesssim 15$\%, and effective rejection of the background 
 of cosmic-ray 
 initiated EAS is achieved under a variety of analyses. An important feature of H.E.S.S. is its $5^\circ$ field-of-view (FoV) 
 diameter, which enables excellent survey coverage of the Galactic plane.

 The SGR\,1806$-$20/Cl*\,1806$-$20 region was covered initially as part of the H.E.S.S. Galactic Plane Survey 
 (HGPS; \citep{hessscanII}, 
 which commenced in 2004.
 Following the tentative indication of a signal, dedicated observation runs were carried out in 2009 and 2010. These runs used 
 the so-called
 wobble mode 
 \citep{hesscrab} in which the region of interest was offset by $0.7^\circ$ from the telescope tracking positions to 
 ensure adequate selection of reflected background regions in 
 spectral analysis. After rejecting observation runs (which are typically 28\,min in duration) based on the presence of clouds and 
 instrumental problems, the 
 total observation time towards SGR\,1806$-$20 amounted to 94\,hours (from approximately 51 and 43\,hours of dedicated and 
 HGPS data, respectively) after correcting for the H.E.S.S. off-axis response and readout dead time.

 In this work we have employed the Model Analysis \citep{hessmodel} (version {\tt HESS\_Soft\_0-8-24}) in which the 
 triggered Cherenkov images from the four 107\,m$^2$ 
 IACTs are compared to a model image.
 Gamma-ray parameters such as arrival direction and energy are then extracted 
 using a log-likelihood maximisation of the differences between the measured and modelled properties.
 The data were analysed using the {\em faint} cuts of the Model Analysis, which employs a minimum image size (total charge) 
 of 120 photoelectrons for the Cherenkov images; similar results were obtained using {\em standard cuts} with a minimum charge 
 of 60 photoelectrons. 
 This analysis was used to generate detection statistics, images, energy
 spectra, and light curves, and has an energy threshold of $E>0.15$\,TeV for observations within $20^\circ$ of the zenith. 
 Averaged over all observations analysed here the threshold is $\sim0.4$\,TeV.
 Consistent results, within statistical and systematic errors, were obtained using 
 alternate background methods such as the {\em template} background \citep{template,berge:2007} and other
 analyses \citep{hesscrab,hessMVA} that employed an independent event calibration procedure.
 Figure~\ref{fig:tevimage} presents the VHE $\gamma$-ray excess count image 
 towards the Cl*\,1806$-$20 region. The {\em ring background} model 
 \citep{berge:2007} was 
 used to estimate the cosmic-ray background. 

 We found that a two-dimensional (2D) symmetrical Gaussian model describes well the intrinsic shape of the source with 
 a 68\% containment radius
 of $0.095^\circ \pm 0.015^\circ$ (or 15\,pc at 8.7\,kpc distance). 
 The fitted position, (J2000.0 epoch), is $\alpha =18^{\rm h}08^{\rm m}37.3^{\rm s}\pm5.1^{\rm s}_{\rm stat}\pm 1.3^{\rm s}_{\rm sys}$ and 
 $\delta =-20^\circ 25^\prime 36.3^{\prime\prime} \pm 71^{\prime\prime}_{\rm stat} \pm 20^{\prime\prime}_{\rm sys}$, 
 with the systematic errors arising from telescope pointing and mechanical alignment uncertainties (see e.g. \citealt{hesscrab}). 
 Based on this we label the source HESS\,J1808$-$204.
 An asymmetric 2D Gaussian model 
 (with major axis, minor axis, position angle anticlockwise from north) = 
 ($0.153^\circ \pm 0.029^\circ$, $0.058^\circ \pm 0.014^\circ$, $50.6^\circ \pm 7.8^\circ$)
 was also well fit to the intrinsic source shape. However, this model was only marginally preferred over a symmetric model 
 (at the 2.4 $\sigma$ level), and so we defaulted to the symmetric model. 

 At the fitted position HESS\,J1808$-$204 has a pre-trial excess significance of $+7.1\sigma$ 
 and comprises 413 gamma-ray photons 
 within a radius of 0.2$^\circ$ (see online Tab~\ref{tab:statistics} for event statistics). This radius is a-priori-selected for 
 extended source searching. After accounting for the 1600 trials in 
 searching for this peak in a $0.4^\circ \times 0.4^\circ$ region around SGR\,1806$-$20 ($40\times40$ bins), the post-trial significance 
 is +6 $\sigma$. Owing to the search binning oversampling the H.E.S.S. analysis point spread function (PSF), and 
 the mixed nature of the data sets (Galactic plane scans and dedicated 
 observations), we consider our post-trial significance to be conservative.
 
\begin{figure}[t]
   \centering
   \includegraphics[bb=20 20 397 283,width=0.52\textwidth]{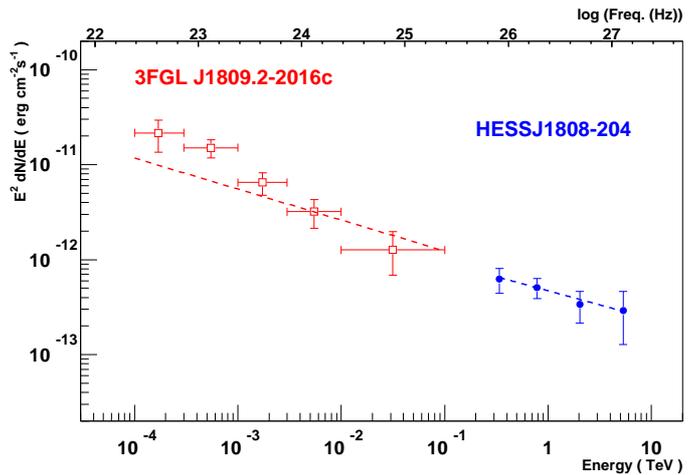}
   \caption{Energy fluxes, 1 $\sigma$ statistical errors, and fitted pure power-law fits for HESS\,J1808$-$204 (blue solid points and blue dashed line) 
     and the {\em Fermi-LAT} source 3FGL\,J1809.2$-$2016c (red open squares and red dashed line) 
     from \citet{fermi3FGL}.}
   \label{fig:spectra}
 \end{figure}

 We calculated the photon spectrum from HESS\,J1808$-$204 
 centred on its fitted position with radius 0.2$^\circ$ to fully encompass the source. The {\em reflected background} 
 model \citep{berge:2007} was used to estimate the cosmic-ray
 background in each energy bin. Table~\ref{tab:spectrum} 
 (online) summarises the photon fluxes and errors.
 The VHE $\gamma$-ray emission was well fit by a power law ($\mathrm{d}N/\mathrm{d}E = \phi (E/{\rm TeV})^{-\Gamma}$) with parameters 
 $\phi = (2.9 \pm 0.4_{\rm stat} \pm 0.5_{\rm sys})\times 10^{-13}$\,ph\,\,cm$^{-2}$\,s$^{-1}$\,TeV$^{-1}$ and  
 $\Gamma = 2.3\pm0.2_{\rm stat}\pm 0.3_{\rm sys}$ (probability = 0.4). 
 The VHE spectral fluxes and power-law fits, and also those of 3FGL\,J1809.2$-$2016c, are shown in Fig.~\ref{fig:spectra} for comparison.

\begin{figure}
   \centering
   \includegraphics[width=0.5\textwidth]{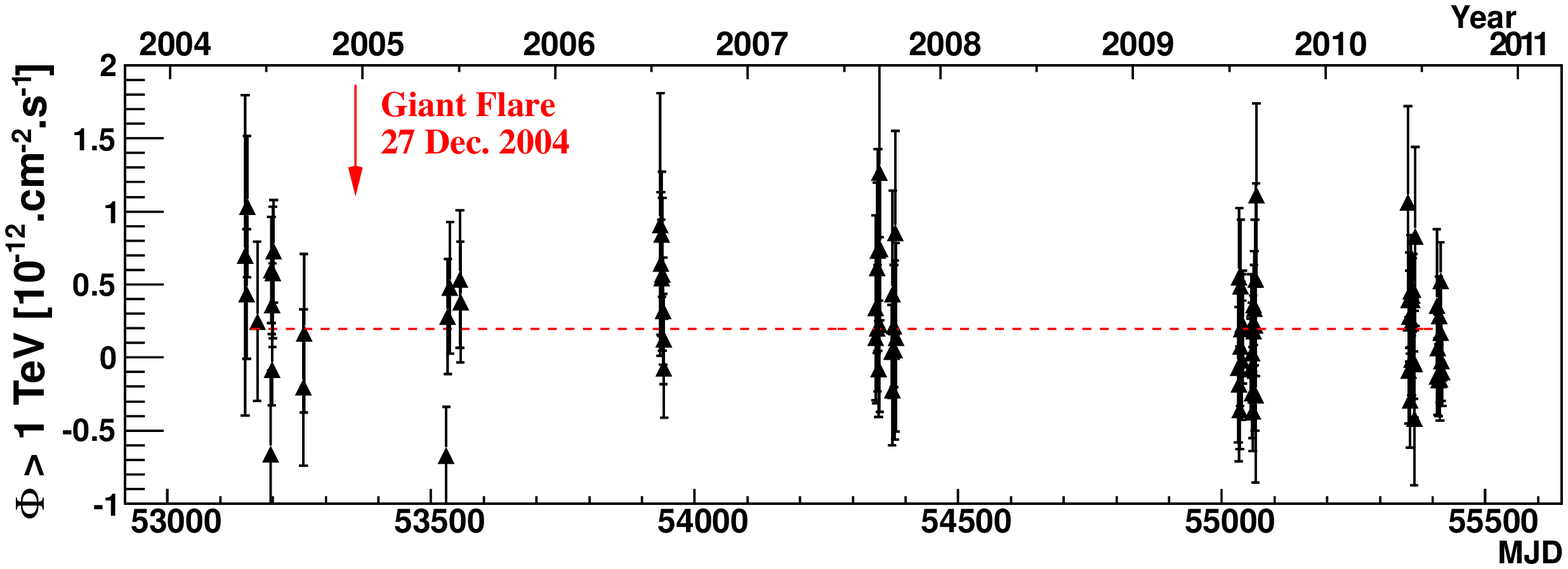} 
   \includegraphics[width=0.5\textwidth]{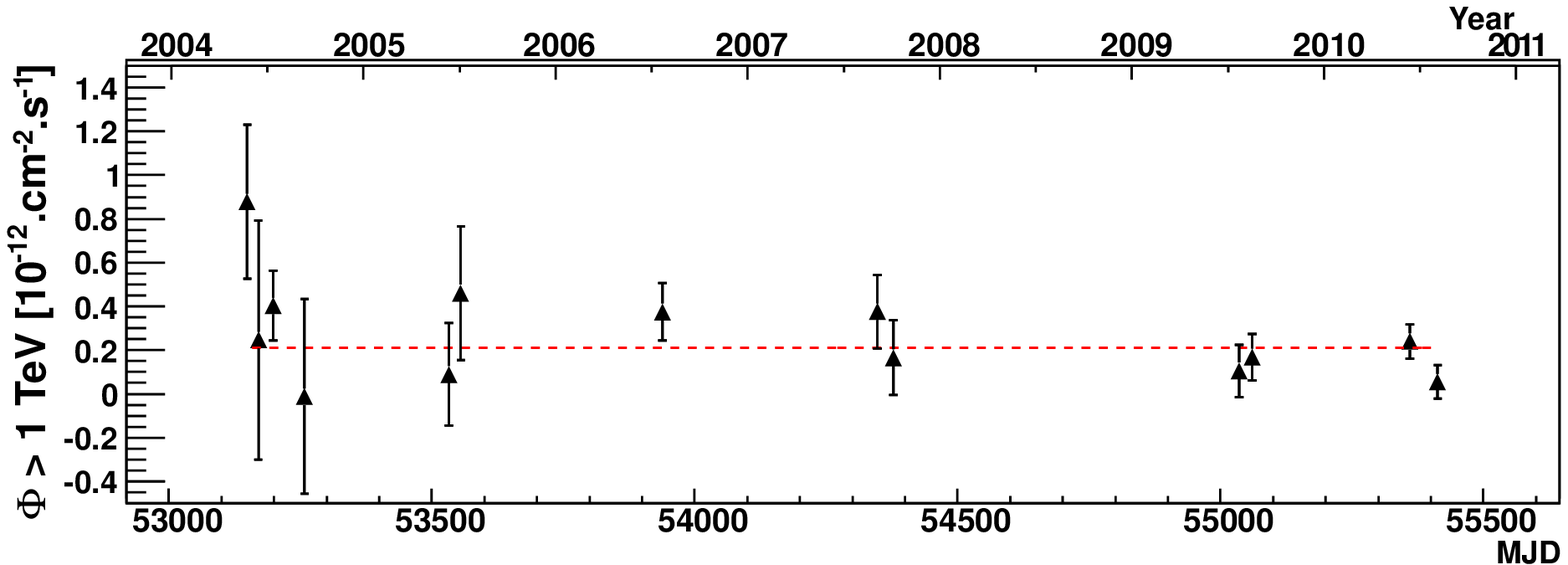} 
   \caption{Light curve of the nightly (top) and lunar monthly (bottom) VHE $\gamma$-ray emission ($\phi>$1\,TeV) 
     from a point-like source (integration radius 0.1$^\circ$) centred on SGR\,1806$-$20. 
     The red dashed line indicates the average flux level in each case and giant flare of 27 December 2004 is indicated on the
     nightly light curve.}
   \label{fig:lc}
 \end{figure}

 Since an additional variable VHE $\gamma$-ray emission component could be expected from SGR\,1806$-$20 and possibly from the member stars of Cl*\,1806$-$20, 
 we examined the flux light curve ($\phi>$1\,TeV) for a point-like test region of radius 0.1$^\circ$, which is optimal for the H.E.S.S. analysis PSF, 
 encompassing these objects over 
 nightly and lunar monthly (dark lunar periods) timescales (see Figure~\ref{fig:lc}).
 We found that the $>$1\,TeV flux light curve was well fit by a steady flux level 
 with $\chi^2/\nu =80.6/88$ for nightly and $\chi^2/\nu =15.2/12$ for lunar monthly timescales, respectively, 
 indicating no evidence for variability in the VHE $\gamma$-ray emission towards SGR\,1806$-$20 and/or Cl*\,1806$-$20.
 A number of soft-gamma-ray flares of SGR\,1806$-$20 occurred since our observations commenced, including the giant flare of 
 27 December 2004, and several other intermediate flares of note (based on announcements from the Gamma-Ray Coordinates Network).
 Our first two periods of H.E.S.S. observations were taken several months on either side of the giant flare. 
 We note that the extent of the H.E.S.S. analysis PSF could 
 integrate steady VHE $\gamma$-ray emission from several sources in the region, 
 diluting any possible variable or periodic emission from a single source such as SGR\,1806$-$20. 
 Additionally, the highly variable spin-down rate $\dot{P}$ of SGR\,1806$-$20 \citep{woods:2007,mereghetti:2011,vigano:2014,younes:2015} 
   over the past decade and infrequently sampled ephemeris (from X-ray measurements) further complicate the search for 
   any pulsed VHE detection. We leave such a study to further work.

\section{Discussion}

\begin{figure}[t]
   \centering
   \includegraphics[width=0.5\textwidth]{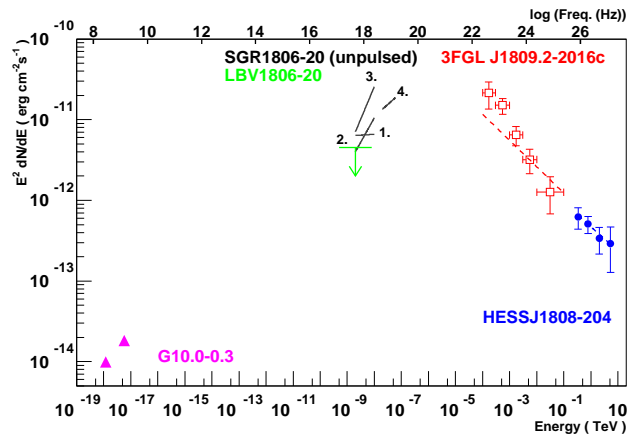}
   \caption{Energy flux (1$\sigma$ statistical error bars) for HESS\,J1801$-$204 and multiwavelength comparisons. 
     Included is the {\em Fermi-LAT} GeV emission \citep{fermi3FGL}, VLA radio emission for G10.0$-$0.3 \citep{kulkarni:1994}, 
     SGR\,1806$-$20 X-ray unpulsed emission (power-law fits: 1. {\em Beppo-SAX} 21 Mar. 1999; 2. {\em XMM-Newton} 3 Apr. 2002; 3. {\em XMM-Newton} 6 Oct. 2004;  and
     4. {\em INTEGRAL IBIS} 2006/2007 from \citet{esposito:2007}), 
     and X-ray upper limit for LBV\,1806$-$20 \citep{naze:2012}.
   }
   \label{fig:sed}
 \end{figure}

\begin{figure}[t]
   \centering
   \includegraphics[width=0.48\textwidth]{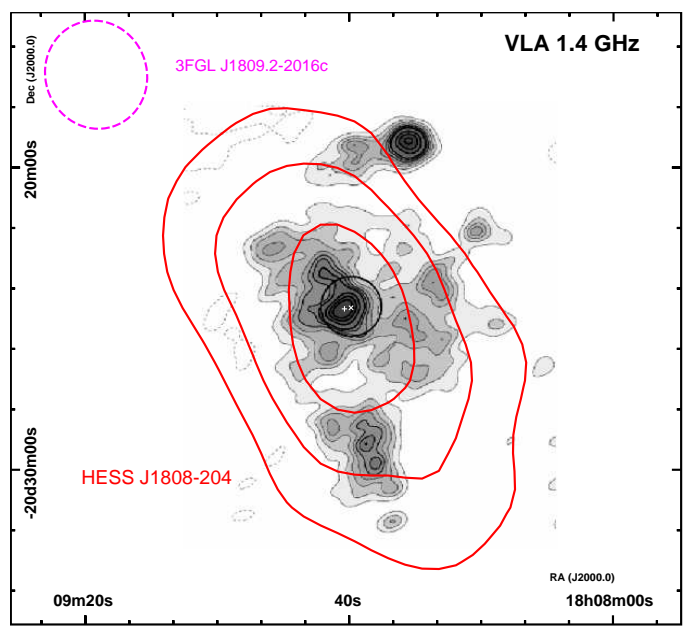}
   \caption{VLA 1.4\,GHz radio image of the radio nebula G10.0$-$0.3 (contour levels at -3, 2, 4, 6, 8, 10, 12, 14, 20, 25, 
     30 $\times$ 1.56 mJy/beam; black circle indicating an {\em ASCA} X-ray source location) from \citet{kulkarni:1994} 
     with red solid VHE $\gamma$-ray contours of HESS\,J1808$-$204 (as per Fig.~\ref{fig:tevimage}).
     The 68\% location error of 3FGL\,J1809.2$-$2016c is indicated by the magenta dashed ellipse, and locations of 
     SGR\,1806$-$20 (x) and LBV\,1806$-$20 (+) are indicated. 
   }
   \label{fig:vla}
 \end{figure}

 Potential counterparts to the VHE $\gamma$-ray source HESS\,J1808$-$204 are the magnetar SGR\,1806$-$20, the massive 
 stellar cluster Cl*\,1806$-$20,
 and/or energetic member stars of the cluster, in particular LBV\,1806$-$20 and/or the WR stars.
 HESS\,J1808$-$204 exhibits an energy flux (0.2 to 10\,TeV) of  
 $F_{\rm VHE}\sim1.7\times 10^{-12}$\,erg\,cm$^{-2}$\,s$^{-1}$ and luminosity of 
 $L_{\rm VHE}\sim1.6 \times 10^{34}(d/{\rm 8.7\, kpc})^2$\,erg\,s$^{-1}$.
 Figure~\ref{fig:sed} presents the energy fluxes of the VHE and multiwavelength sources in this region (highlighting the prominence of the
 $\gamma$-ray emission), including an X-ray upper limit 
 for LBV\,1806$-$20 \citep{naze:2012}. There are no signs of any SNRs (although one should exist to explain SGR\,1806$-$20) 
 or other energetic pulsars in the region. However, a prominent multiwavelength feature
 is the synchrotron radio nebula G10.0$-$0.3. 
 Interestingly, HESS\,J1808$-$204 is very similar in intrinsic size to 
 G10.0$-$0.3 as shown in Fig.~\ref{fig:vla}.  
 Based on this, LBV\,1806$-$20 (the possible source of energy for G10.0$-$0.3) could be considered a plausible counterpart.

 With a likely kinetic stellar wind luminosity of $L_{\rm W}>10^{38}$\,erg\,s$^{-1}$, the cluster Cl*\,1806$-$20 could 
 easily account for the VHE $\gamma$-ray luminosity of HESS\,J1808$-$204 and also of the nearby {\em Fermi-LAT} source. 
 Of the member stars, the four WR stars and in particular LBV\,1806$-$20 could dominate this stellar wind energy and therefore 
 drive much of the particle acceleration. The VHE $\gamma$-ray emission would represent only a small fraction $\sim\,10^{-4}$ of the 
 cLBV's wind luminosity, L$_{\rm LBV}\sim10^{38}$\,erg\,s$^{-1}$, for example.
 
 In a scenario involving the cluster of massive stars, particle acceleration could take place as a result of the stellar wind 
 interaction over parsec scales according to the cluster size and stellar density. 
 Several extended VHE $\gamma$-ray sources have already been linked to such processes likely to be present in massive 
 stellar clusters (e.g. in Cyg\,OB2, Westerlund\,2 and Westerlund\,1; \citealt{tevj2032,hesswd2,hesswd1}). Here, Westerlund\,1
 may bear some resemblance to Cl*\,1806$-$20 as it also harbours a magnetar (the anomalous X-ray pulsar 
 CXOU\,J164710.2$-$455216; \citealt{muno:2007}) 
 and a LBV star (Wd1-243; see e.g. \citealt{clark:2008}). A point of difference is that the VHE emission towards Westerlund\,1 is 
 physically about a factor 10 larger than that towards Cl*\,1806$-$20 (radii of 160\,pc vs. 15\,pc).
  
 The luminosity of LBV\,1806$-$20, $L\sim10^{6.3}$L$_\odot$, ranks it in the top five amongst the 35 LBV and cLBV  
 list \citep{clark:2005} (see \citealt{kniazev:2015} for the most recent catalogue of LBVs).  
 VHE $\gamma$-ray emission is found towards four of the top five in this list: the {\em Pistol-Star} and FMM\,362 towards the 
 diffuse VHE Galactic centre emission \citep{hessgalcen}, Cyg\,OB2\#12 towards TeV\,J2032+4130 \citep{tevj2032}, and Wray\,17-96 
 in the vicinity of HESS\,J1741$-$302 \citep{tibolla:2009}. It remains to be seen, however, how these LBVs are related to their nearby
 VHE $\gamma$-ray sources. The notable exception is $\eta$-Car, which is probably the most luminous LBV.  
 $\eta$-Car exhibits $\gamma$-ray emission (up to about 200\,GeV) modulated by its orbital period of 5.54\,yr
 \citep{etacarHESS,etacarfermi,etacaragile,reitberger:2015}, suggesting a wind-wind interaction as the source of particle
 acceleration. A similar process may be active in LBV\,1806$-$20 since it too may be a binary system.
 
 The linkage between HESS\,J1808$-$204 and the {\em Fermi-LAT} GeV source 3FGL\,J1809.2$-$2016c is not 
 so clear given the significant spatial
 separation between the two and the confused nature of the GeV source. For confused GeV sources there are considerable systematic 
   uncertainties in location and existence above the diffuse GeV background. As a result, the sub-GeV spectral determination
   is usually compromised (as is readily apparent in Figs.~\ref{fig:spectra} and \ref{fig:sed}).
 Thus, even though the fitted power -law 
 index of the GeV source ($\Gamma_{\rm GeV} = 2.33\pm0.09$ from \citealt{fermi3FGL}) is very similar to 
 that of HESS\,J1808$-$204,
 and its extrapolation to TeV energies is within systematic errors that are consistent with the VHE flux points,
 the above-mentioned caveats prevent further detailed comparisons. In the case in which the GeV and VHE sources are 
 nevertheless connected via the same putative particle accelerator, 
 the $\gamma$-ray luminosity would increase by 
 about a factor of 5 to 10 to account for the dominant 
 GeV component ($L_{\rm GeV}\sim10^{35}$\,erg\,s$^{-1}$ at 8.7\,kpc).
  
 Turning our attention to SGR\,1806$-$20, the VHE $\gamma$-ray luminosity is up to $\sim$50\% of this magnetar's spin-down power. 
 The established VHE pulsar wind nebulae (PWN) have VHE spin-down 
 efficiencies of $L_{\rm VHE} / L_{\rm SD} \lesssim $10\% and are associated with pulsars exhibiting high spin-down power of 
 $L_{\rm SD}>10^{35}$\,erg\,s$^{-1}$ \citep{halpern:2010}, in contrast to the situation here. The quiescent X-ray luminosity for 
 SGR\,1806$-$20 is also similar to or greater than its spin-down power, 
 suggesting the X-rays result from magnetic energy perhaps via reconnection \citep{zhang:2003} rather than the rotational 
 energy associated with pulsar spin-down. The luminosity of the nearby {\em Fermi-LAT} GeV source could also be met by magnetic energy. 
 Moreover, the rather low ratio $L_{\rm VHE}/L_{\rm X}\sim0.1$ (for the X-ray luminosity 
 $L_{\rm X}$ in the 2 to 10\,keV range) is clearly at odds with the 
 observed trend of $L_{\rm VHE}/L_{\rm X} \ga 10$ for pulsars with  $L_{\rm SD}<10^{36}$\,erg\,s$^{-1}$ \citep{mattana:2009}. 
 Hence if HESS\,J1808$-$240 is attributed to electrons accelerated by SGR\,1806$-$20, it is likely that magnetic energy is the 
 source of power. 

 Given the presence of several molecular clouds along the line of sight towards Cl*\,1806$-$20  \citep{corbel:2004}, 
 a hadronic origin for 
 HESS\,J1808$-$204 involving collisions of multi-TeV protons with interstellar gas is worth considering.
 The singly pointed CO observations of \citet{corbel:2004} however do not reveal the spatial distribution, mass, or density of 
 the molecular clouds. Studies of the molecular cloud morphology 
 are hence needed to ascertain the detailed likelihood of a hadronic origin for the HESS\,J1808$-$204 and the transport properties of
 particles from the Cl*\,1806$-$20 region.
 Nevertheless based on the integrated CO brightness temperatures reported in Table 2 of \citet{corbel:2004}, we estimated approximate 
 proton densities $n$ of the order of 10$^3$\,cm$^{-3}$ for the brighter clouds MC\,13A, MC\,73, and MC\,44 (as labelled by
  \citealt{corbel:2004}) at their respective distances when integrating
 over the 45$^{\prime\prime}$ beam full width at half maximum (FWHM) of the telescope used in the CO observations (\citealt{corbel:2004} 
 favoured MC\,13A as most likely
 associated with the cluster). These densities may be considered upper
 limits to the wider cloud densities since the CO measurements were taken towards the stellar cluster where molecular gas 
 density may be expected to peak. 

 Using the relation $B\sim10(n\,/\,300 {\rm cm}^{-3})^{0.65}\mu$G from \citet{crutcher:2010}, 
 magnetic fields of order $B\sim20\mu$G
 could be expected inside the clouds.
 From here and assuming a turbulent B-field, we can employ the formalism of \citet{gabici:2007} (Eq.\,2) for the diffusion 
 coefficient as a function
 of magnetic field $B$ and suppression factor $\chi$, which is used to account for increased 
 magnetic field turbulence related to cosmic-ray streaming instabilities in the interstellar medium. We note however that
 the cosmic-ray transport may not necessarily be diffusive in the case of a more ordered B-field structure, as for example indicated
 by \citet{crutcher:2010} in molecular clouds of low density $n<300$\,cm$^{-3}$.
 We can, nevertheless, use the young age (650\,years) of SGR\,1806$-$20 to provide some limits on the diffusion distances of particles. 
 For 50\,TeV protons coming from the magnetar and assuming $\chi=1$,
 we arrive at a diffusion 
 length of $l\sim$\,30\,pc, which is approaching the observed $\sim$\,15\,pc radius of HESS\,J1808$-$204. The parameter $\chi$ is poorly 
 constrained, but
 \citet{protheroe:2008} have suggested $\chi<0.01 - 0.1$  based on observations of the GeV $\gamma$-ray emission towards the dense 
 Sgr\,B2 giant molecular cloud.  

 We can also infer the cosmic-ray proton energy budget $W_p = L_{\rm VHE} {\rm (erg\,s^{-1})} \tau_{\rm pp}$\,erg required to power 
 HESS\,J1808$-$204. Here,
 $\tau_{\rm pp}\sim1.7\times 10^{15} / n\,{\rm cm^{-3}}$\,s is the cooling time for $\gamma$-rays produced by proton-proton collisions
 as a function of the target density $n$.
 For the density discussed above we find $W_p\sim10^{46}$\,erg. This is rather modest compared to that of a canonical
 supernova remnant, however, our estimate for $W_p$ would be a lower
limit since the density is likely
 an upper limit as explained above. 

 An alternative, leptonic origin for HESS\,J1808$-$204 might arise from IC scattering of local soft photon fields by TeV 
 to multi-TeV electrons.
 The local infrared (IR) field (due to heated dust) peaks strongly towards Cl*\,1806$-$20 \citep{rahoui:2009} with an energy density of 
 $\sim$\,20 to 50\,eV\,cm$^{-3}$ as measured by {\em Spitzer} at 24 $\mu$m within $\sim30^{\prime\prime}$ of Cl*\,1806$-$20 
 (see online Fig.~\ref{fig:ir_image}). 
 Measured over the 15\,pc radius of VHE $\gamma$-ray emission region, however, the IR energy density, $\sim$\,0.4\,eV\,cm$^{-3}$, is 
 comparable to that of the cosmic microwave background (CMB) at 0.25\,eV\,cm$^{-3}$. An additional soft photon field 
 comes from the optical/UV photons
 from the massive stellar content of Cl*\,1806$-$20. Based on the presence of five OB, four WR stars, and one 
 cLBV
 \citep{figer:2005,edwards:2011}, we estimated a bolometric luminosity of $\sim\,10^{40}$\,erg\,s$^{-1}$. Averaged over the 
 VHE $\gamma$-ray emission region, the resulting optical/UV energy density is $\sim$\,7\,eV\,cm$^{-3}$.

 Considering the TeV to multi-TeV electron energy loss rate for IC scattering taking into account Klein-Nishina effects 
 (e.g. Eq. 35 of \citealt{aharonian:1981}) with these energy densities, 
 we find that the dominant IC component will likely come from up-scattered CMB photons in the Thomson scattering regime 
 as the Klein-Nishina effect will suppress the IR and optical/UV components. Thus we can take the X-ray power-law 
 components of SGR\,1806$-$20 ($F_X\sim{\rm few}\,\times10^{-11}$erg\,s$^{-1}$\,cm$^{-2}$) as synchrotron emission 
 from multi-TeV electrons 
 and the VHE $\gamma$-ray flux $F_{\rm VHE}$
 as arising from IC scattering of the CMB photons by the same electrons. 
 From consideration of the IC and synchrotron luminosities, assuming IC emission only comes from up-scattered CMB 
   photons in the Thomson regime, the magnetic 
 field $B\sim10\sqrt{\xi F_X / (10 F_{\rm VHE})}\,\mu$G in the region common to both fluxes can be estimated. 
 Here the factor $\xi$ accounts for the radii of the VHE and X-ray emission ($\xi=(R_{\rm VHE}/R_{\rm X})^2$). 
 
 We estimate $B\gtrsim1$\,mG for the values 
 $R_{\rm VHE}\sim360^{\prime \prime}$ (VHE radius) and $R_{\rm X}\lesssim3^{\prime
\prime}$ \citep{kaplan:2002}, which might be expected
 given the extreme magnetic field of SGR\,1806$-$20 and that of the massive stars in Cl*\,1806$-$20.
 The sub-parsec size of the X-ray emission region \citep{kaplan:2002} is also consistent with a magnetic field 
 $B\sim$\,few\,mG if the synchrotron cooling time ($t_{\rm sync}<$\,few\,yr) dominantly limits the transport of the parent multi-TeV 
 electrons.
 There are in fact a number of other high magnetic field pulsars with compact X-ray nebulae 
 potentially associated with unidentified VHE $\gamma$-ray sources (e.g. see \citealt{kargaltsev:2013}) that may be explained 
 within this scenario.

 The 15\,pc radius of the TeV emission region could be allowed by the much longer IC cooling time of $t_{\rm IC}\sim10^{3}-10^{4}$\,yr, 
 provided that the magnetic field has declined to $\lesssim10 \mu$G outside the X-ray region to avoid synchrotron losses.
 A reduced $B$ field is in fact implied by the X-ray upper limit ($4.53\times10^{-12}$erg\,s$^{-1}$\,cm$^{-2}$) for 
 LBV\,1806$-$20 \citep{naze:2012}, and is only 15$^{\prime\prime}$ away from SGR\,1806$-$20.  Such a reduction in the
 B-field may also play a role in limiting the X-ray emission region size around SGR\,1806$-$20, by permitting electrons
 to escape to the wider IC-dominated region; the electron diffusion coefficient would likely increase as well, 
 further enhancing their escape. 
 Our inferred B-field value of $\sim20 \mu$G from the molecular cloud column density as discussed earlier, may still
 limit the level of IC emission. As argued earlier for the hadronic interpretation, however, spatial studies of the 
 molecular gas will be needed to more confidently discriminate hadronic and leptonic models for HESS\,J1808$-$204.
 
 \section{Conclusions}
 We report the discovery with the H.E.S.S. telescopes of extended VHE $\gamma$-ray emission (HESS\,J1808$-$204) 
 towards the luminous blue variable candidate LBV\,1806$-$20, the massive stellar cluster Cl*\,1806$-$20, and 
   the magnetar SGR\,1806$-$20. The H.E.S.S.
 telescopes are not able to resolve these potential counterparts, which are located within 
   a 0.5$^\prime$ radius. However the extension of the $\gamma$-ray emission, 
 at $\sim\,0.1^\circ$ radius  (or 15\,pc for a distance of 8.7\,kpc) is similar in scale to the radio nebula G10.0$-$0.3 
 supposedly powered by LBV\,1806$-$20. 
 The intense stellar wind luminosity of LBV\,1806$-$20, by itself or collectively from the other massive stars in 
 Cl*\,1806$-$20, could readily power the VHE source, which has a luminosity of 
 $L_{\rm VHE}\sim1.6 \times 10^{34}(d/{\rm 8.7\, kpc})^2$\,erg\,s$^{-1}$. If associated with SGR\,1806$-$20, the reported 
 young age of 650\,yr for this magnetar, 
 along with our estimated magnetic field of $20 \mu$G, could  
 imply a diffusive transport limit of $<$30\,pc which is similar to the size of the VHE emission. 
 Additionally, in this case the VHE luminosity could only realistically be met by magnetic dissipative effects rather than 
 the magnetar spin-down process. Whatever the origin of the parent particles responsible for 
 HESS\,J1804$-$204, their hadronic and/or leptonic nature is currently unclear.
 Detailed observations of the molecular gas spatial distribution would be needed for some discrimination of 
 hadronic from leptonic scenarios (in particular by providing constraints on the magnetic field in the region). 
 Looking towards the future, the arc-minute angular resolution 
 of the forthcoming Cherenkov Telescope Array \citep{CTA} will be  
 able to probe the VHE morphology on the parsec scales necessary to probe for energy dependent morphology, providing 
 further information about the nature and origin of the particles responsible for HESS\,J1804$-$204.
 In summary, the discovery of HESS\,J1808$-$204 provides further impetus to the notion that magnetars, 
 massive stars, and/or stellar clusters can accelerate particles to beyond TeV energies. 

 \begin{acknowledgements}
   The support of the Namibian authorities and of the University of Namibia in facilitating the construction and operation of H.E.S.S. 
   is gratefully acknowledged, as is the support by the German Ministry for Education and Research (BMBF), the Max Planck Society, the 
   German Research Foundation (DFG), the French Ministry for Research, the CNRS-IN2P3 and the Astroparticle Interdisciplinary Programme 
   of the CNRS, the U.K. Science and Technology Facilities Council (STFC), the IPNP of the Charles University, the Czech Science 
   Foundation, the Polish Ministry of Science and Higher Education, the South African Department of Science and Technology and National 
   Research Foundation, the Australian Research Council and by the University of Namibia. We appreciate the excellent work of 
   the technical support staff in Berlin, Durham, Hamburg, Heidelberg, Palaiseau, Paris, Saclay, and in Namibia in the construction and 
   operation of the equipment.
 \end{acknowledgements}

\bibliography{28695_am-gpr}{}

\begin{thebibliography}{75}
\expandafter\ifx\csname natexlab\endcsname\relax\def\natexlab#1{#1}\fi

\bibitem[{{Abdo} {et~al.}(2010){Abdo}, {Ackermann}, {Ajello}, {Allafort},
  {Baldini}, {Ballet}, {Barbiellini}, {Bastieri}, {Bechtol}, {Bellazzini},
  {Berenji}, {Blandford}, {Bonamente}, {Borgland}, {Bouvier}, {Brandt},
  {Bregeon}, {Brez}, {Brigida}, {Bruel}, {Buehler}, {Burnett}, {Caliandro},
  {Cameron}, {Caraveo}, {Carrigan}, {Casandjian}, {Cecchi}, {{\c C}elik},
  {Chaty}, {Chekhtman}, {Cheung}, {Chiang}, {Ciprini}, {Claus}, {Cohen-Tanugi},
  {Cominsky}, {Conrad}, {Dermer}, {de Palma}, {Digel}, {Silva}, {Drell},
  {Dubois}, {Dumora}, {Favuzzi}, {Fegan}, {Ferrara}, {Frailis}, {Fukazawa},
  {Fusco}, {Gargano}, {Gehrels}, {Germani}, {Giglietto}, {Giordano}, {Godfrey},
  {Grenier}, {Grondin}, {Grove}, {Guillemot}, {Guiriec}, {Hadasch}, {Hanabata},
  {Harding}, {Hayashida}, {Hays}, {Hill}, {Horan}, {Hughes}, {Itoh}, {Jackson},
  {J{\'o}hannesson}, {Johnson}, {Johnson}, {Kamae}, {Katagiri}, {Kataoka},
  {Kerr}, {Kn{\"o}dlseder}, {Kuss}, {Lande}, {Latronico}, {Lee},
  {Lemoine-Goumard}, {Livingstone}, {Llena Garde}, {Longo}, {Loparco},
  {Lovellette}, {Lubrano}, {Makeev}, {Mazziotta}, {McEnery}, {Mehault},
  {Michelson}, {Mitthumsiri}, {Mizuno}, {Moiseev}, {Monte}, {Monzani},
  {Morselli}, {Moskalenko}, {Murgia}, {Nakamori}, {Naumann-Godo}, {Nolan},
  {Norris}, {Nuss}, {Ohsugi}, {Okumura}, {Omodei}, {Orlando}, {Ormes}, {Ozaki},
  {Panetta}, {Parent}, {Pelassa}, {Pepe}, {Pesce-Rollins}, {Piron}, {Porter},
  {Rain{\`o}}, {Rando}, {Razzano}, {Reimer}, {Reimer}, {Reposeur}, {Rodriguez},
  {Romani}, {Roth}, {Sadrozinski}, {Sander}, {Saz Parkinson}, {Scargle},
  {Sgr{\`o}}, {Siskind}, {Smith}, {Smith}, {Spandre}, {Spinelli}, {Strickman},
  {Suson}, {Takahashi}, {Takahashi}, {Tanaka}, {Thayer}, {Thayer}, {Thompson},
  {Tibaldo}, {Tibolla}, {Torres}, {Tosti}, {Tramacere}, {Uchiyama}, {Usher},
  {Vandenbroucke}, {Vasileiou}, {Vilchez}, {Vitale}, {Waite}, {Wallace},
  {Wang}, {Winer}, {Wood}, {Yang}, {Ylinen}, \& {Ziegler}}]{etacarfermi}
{Abdo}, A.~A., {Ackermann}, M., {Ajello}, M., {et~al.} 2010, \apj, 723, 649

\bibitem[{{Abramowski} {et~al.}(2012{\natexlab{a}}){Abramowski}, {Acero},
  {Aharonian}, {Akhperjanian}, {Anton}, {Balzer}, {Barnacka}, {Barres de
  Almeida}, {Becherini}, {Becker}, {Behera}, {Bernl{\"o}hr}, {Birsin},
  {Biteau}, {Bochow}, {Boisson}, {Bolmont}, {Bordas}, {Brucker}, {Brun},
  {Brun}, {Bulik}, {B{\"u}sching}, {Carrigan}, {Casanova}, {Cerruti},
  {Chadwick}, {Charbonnier}, {Chaves}, {Cheesebrough}, {Chounet}, {Clapson},
  {Coignet}, {Cologna}, {Conrad}, {Dalton}, {Daniel}, {Davids}, {Degrange},
  {Deil}, {Dickinson}, {Djannati-Ata{\"i}}, {Domainko}, {O'C.~Drury}, {Dubois},
  {Dubus}, {Dutson}, {Dyks}, {Dyrda}, {Egberts}, {Eger}, {Espigat}, {Fallon},
  {Farnier}, {Fegan}, {Feinstein}, {Fernandes}, {Fiasson}, {Fontaine},
  {F{\"o}rster}, {F{\"u}{\ss}ling}, {Gallant}, {Gast}, {G{\'e}rard}, {Gerbig},
  {Giebels}, {Glicenstein}, {Gl{\"u}ck}, {Goret}, {G{\"o}ring}, {H{\"a}ffner},
  {Hague}, {Hampf}, {Hauser}, {Heinz}, {Heinzelmann}, {Henri}, {Hermann},
  {Hinton}, {Hoffmann}, {Hofmann}, {Hofverberg}, {Holler}, {Horns},
  {Jacholkowska}, {de Jager}, {Jahn}, {Jamrozy}, {Jung}, {Kastendieck},
  {Katarzy{\'n}ski}, {Katz}, {Kaufmann}, {Keogh}, {Khangulyan}, {Kh{\'e}lifi},
  {Klochkov}, {Klu{\.z}niak}, {Kneiske}, {Komin}, {Kosack}, {Kossakowski},
  {Laffon}, {Lamanna}, {Lennarz}, {Lohse}, {Lopatin}, {Lu}, {Marandon},
  {Marcowith}, {Masbou}, {Maurin}, {Maxted}, {Mayer}, {McComb}, {Medina},
  {M{\'e}hault}, {Moderski}, {Moulin}, {Naumann}, {Naumann-Godo}, {de Naurois},
  {Nedbal}, {Nekrassov}, {Nguyen}, {Nicholas}, {Niemiec}, {Nolan}, {Ohm}, {de
  O{\~n}a Wilhelmi}, {Opitz}, {Ostrowski}, {Oya}, {Panter}, {Paz Arribas},
  {Pedaletti}, {Pelletier}, {Petrucci}, {Pita}, {P{\"u}hlhofer}, {Punch},
  {Quirrenbach}, {Raue}, {Rayner}, {Reimer}, {Reimer}, {Renaud}, {de Los
  Reyes}, {Rieger}, {Ripken}, {Rob}, {Rosier-Lees}, {Rowell}, {Rudak},
  {Rulten}, {Ruppel}, {Sahakian}, {Sanchez}, {Santangelo}, {Schlickeiser},
  {Sch{\"o}ck}, {Schulz}, {Schwanke}, {Schwarzburg}, {Schwemmer}, {Sheidaei},
  {Sikora}, {Skilton}, {Sol}, {Spengler}, {Stawarz}, {Steenkamp}, {Stegmann},
  {Stinzing}, {Stycz}, {Sushch}, {Szostek}, {Tavernet}, {Terrier},
  {Tluczykont}, {Valerius}, {van Eldik}, {Vasileiadis}, {Venter}, {Vialle},
  {Viana}, {Vincent}, {V{\"o}lk}, {Volpe}, {Vorobiov}, {Vorster}, {Wagner},
  {Ward}, {White}, {Wierzcholska}, {Zacharias}, {Zajczyk}, {Zdziarski}, {Zech},
  \& {Zechlin}}]{hesswd1}
{Abramowski}, A., {Acero}, F., {Aharonian}, F., {et~al.} 2012{\natexlab{a}},
  \aap, 537, A114

\bibitem[{{Abramowski} {et~al.}(2012{\natexlab{b}}){Abramowski}, {Acero},
  {Aharonian}, {Akhperjanian}, {Anton}, {Balzer}, {Barnacka}, {Becherini},
  {Becker}, {Bernl{\"o}hr}, {Birsin}, {Biteau}, {Bochow}, {Boisson}, {Bolmont},
  {Bordas}, {Brucker}, {Brun}, {Brun}, {Bulik}, {B{\"u}sching}, {Carrigan},
  {Casanova}, {Cerruti}, {Chadwick}, {Charbonnier}, {Chaves}, {Cheesebrough},
  {Cologna}, {Conrad}, {Dalton}, {Daniel}, {Davids}, {Degrange}, {Deil},
  {Dickinson}, {Djannati-Ata{\"i}}, {Domainko}, {Drury}, {Dubus}, {Dutson},
  {Dyks}, {Dyrda}, {Egberts}, {Eger}, {Espigat}, {Fallon}, {Fegan},
  {Feinstein}, {Fernandes}, {Fiasson}, {Fontaine}, {F{\"o}rster},
  {F{\"u}{\ss}ling}, {Gallant}, {Garrigoux}, {Gast}, {G{\'e}rard}, {Giebels},
  {Glicenstein}, {Gl{\"u}ck}, {G{\"o}ring}, {Grondin}, {H{\"a}ffner}, {Hague},
  {Hahn}, {Hampf}, {Harris}, {Hauser}, {Heinz}, {Heinzelmann}, {Henri},
  {Hermann}, {Hillert}, {Hinton}, {Hofmann}, {Hofverberg}, {Holler}, {Horns},
  {Jacholkowska}, {Jahn}, {Jamrozy}, {Jung}, {Kastendieck}, {Katarzy{\'n}ski},
  {Katz}, {Kaufmann}, {Kh{\'e}lifi}, {Klochkov}, {Klu{\'z}niak}, {Kneiske},
  {Komin}, {Kosack}, {Kossakowski}, {Krayzel}, {Laffon}, {Lamanna}, {Lenain},
  {Lennarz}, {Lohse}, {Lopatin}, {Lu}, {Marandon}, {Marcowith}, {Masbou},
  {Maurin}, {Maxted}, {Mayer}, {McComb}, {Medina}, {M{\'e}hault}, {Moderski},
  {Mohamed}, {Moulin}, {Naumann}, {Naumann-Godo}, {de Naurois}, {Nedbal},
  {Nekrassov}, {Nguyen}, {Nicholas}, {Niemiec}, {Nolan}, {Ohm}, {de O{\~n}a
  Wilhelmi}, {Opitz}, {Ostrowski}, {Oya}, {Panter}, {Paz Arribas}, {Pekeur},
  {Pelletier}, {Perez}, {Petrucci}, {Peyaud}, {Pita}, {P{\"u}hlhofer}, {Punch},
  {Quirrenbach}, {Raue}, {Reimer}, {Reimer}, {Renaud}, {de los Reyes},
  {Rieger}, {Ripken}, {Rob}, {Rosier-Lees}, {Rowell}, {Rudak}, {Rulten},
  {Sahakian}, {Sanchez}, {Santangelo}, {Schlickeiser}, {Schulz}, {Schwanke},
  {Schwarzburg}, {Schwemmer}, {Sheidaei}, {Skilton}, {Sol}, {Spengler},
  {Stawarz}, {Steenkamp}, {Stegmann}, {Stinzing}, {Stycz}, {Sushch}, {Szostek},
  {Tavernet}, {Terrier}, {Tluczykont}, {Valerius}, {van Eldik}, {Vasileiadis},
  {Venter}, {Viana}, {Vincent}, {V{\"o}lk}, {Volpe}, {Vorobiov}, {Vorster},
  {Wagner}, {Ward}, {White}, {Wierzcholska}, {Zacharias}, {Zajczyk},
  {Zdziarski}, {Zech}, {Zechlin}, \& {Montmerle}}]{etacarHESS}
{Abramowski}, A., {Acero}, F., {Aharonian}, F., {et~al.} 2012{\natexlab{b}},
  \mnras, 424, 128

\bibitem[{{Acero} {et~al.}(2015){Acero}, {Ackermann}, {Ajello}, {Albert},
  {Atwood}, {Axelsson}, {Baldini}, {Ballet}, {Barbiellini}, {Bastieri},
  {Belfiore}, {Bellazzini}, {Bissaldi}, {Blandford}, {Bloom}, {Bogart},
  {Bonino}, {Bottacini}, {Bregeon}, {Britto}, {Bruel}, {Buehler}, {Burnett},
  {Buson}, {Caliandro}, {Cameron}, {Caputo}, {Caragiulo}, {Caraveo},
  {Casandjian}, {Cavazzuti}, {Charles}, {Chaves}, {Chekhtman}, {Cheung},
  {Chiang}, {Chiaro}, {Ciprini}, {Claus}, {Cohen-Tanugi}, {Cominsky}, {Conrad},
  {Cutini}, {D{\'}Ammando}, {de Angelis}, {DeKlotz}, {de Palma}, {Desiante},
  {Digel}, {Di Venere}, {Drell}, {Dubois}, {Dumora}, {Favuzzi}, {Fegan},
  {Ferrara}, {Finke}, {Franckowiak}, {Fukazawa}, {Funk}, {Fusco}, {Gargano},
  {Gasparrini}, {Giebels}, {Giglietto}, {Giommi}, {Giordano}, {Giroletti},
  {Glanzman}, {Godfrey}, {Grenier}, {Grondin}, {Grove}, {Guillemot}, {Guiriec},
  {Hadasch}, {Harding}, {Hays}, {Hewitt}, {Hill}, {Horan}, {Iafrate}, {Jogler},
  {J{\'o}hannesson}, {Johnson}, {Johnson}, {Johnson}, {Johnson}, {Kamae},
  {Kataoka}, {Katsuta}, {Kuss}, {La Mura}, {Landriu}, {Larsson}, {Latronico},
  {Lemoine-Goumard}, {Li}, {Li}, {Longo}, {Loparco}, {Lott}, {Lovellette},
  {Lubrano}, {Madejski}, {Massaro}, {Mayer}, {Mazziotta}, {McEnery},
  {Michelson}, {Mirabal}, {Mizuno}, {Moiseev}, {Mongelli}, {Monzani},
  {Morselli}, {Moskalenko}, {Murgia}, {Nuss}, {Ohno}, {Ohsugi}, {Omodei},
  {Orienti}, {Orlando}, {Ormes}, {Paneque}, {Panetta}, {Perkins},
  {Pesce-Rollins}, {Piron}, {Pivato}, {Porter}, {Racusin}, {Rando}, {Razzano},
  {Razzaque}, {Reimer}, {Reimer}, {Reposeur}, {Rochester}, {Romani},
  {Salvetti}, {S{\'a}nchez-Conde}, {Saz Parkinson}, {Schulz}, {Siskind},
  {Smith}, {Spada}, {Spandre}, {Spinelli}, {Stephens}, {Strong}, {Suson},
  {Takahashi}, {Takahashi}, {Tanaka}, {Thayer}, {Thayer}, {Thompson},
  {Tibaldo}, {Tibolla}, {Torres}, {Torresi}, {Tosti}, {Troja}, {Van Klaveren},
  {Vianello}, {Winer}, {Wood}, {Wood}, \& {Zimmer}}]{fermi3FGL}
{Acero}, F., {Ackermann}, M., {Ajello}, M., {et~al.} 2015, \apjs, 218, 23

\bibitem[{{Acharya} {et~al.}(2013){Acharya}, {Actis}, {Aghajani}, {Agnetta},
  {Aguilar}, {Aharonian}, {Ajello}, {Akhperjanian}, {Alcubierre},
  {Aleksi{\'c}}, \& et~al.}]{CTA}
{Acharya}, B.~S., {Actis}, M., {Aghajani}, T., {et~al.} 2013, Astroparticle
  Physics, 43, 3

\bibitem[{{Aharonian} {et~al.}(2002){Aharonian}, {Akhperjanian}, {Beilicke},
  {Bernl{\"o}hr}, {B{\"o}rst}, {Bojahr}, {Bolz}, {Coarasa}, {Contreras},
  {Cortina}, {Denninghoff}, {Fonseca}, {Girma}, {G{\"o}tting}, {Heinzelmann},
  {Hermann}, {Heusler}, {Hofmann}, {Horns}, {Jung}, {Kankanyan}, {Kestel},
  {Kettler}, {Kohnle}, {Konopelko}, {Kornmeyer}, {Kranich}, {Krawczynski},
  {Lampeitl}, {Lopez}, {Lorenz}, {Lucarelli}, {Magnussen}, {Mang}, {Meyer},
  {Milite}, {Mirzoyan}, {Moralejo}, {Ona}, {Panter}, {Plyasheshnikov}, {Prahl},
  {P{\"u}hlhofer}, {Rauterberg}, {Reyes}, {Rhode}, {Ripken}, {R{\"o}hring},
  {Rowell}, {Sahakian}, {Samorski}, {Schilling}, {Schr{\"o}der}, {Siems},
  {Sobzynska}, {Stamm}, {Tluczykont}, {V{\"o}lk}, {Wiedner}, {Wittek},
  {Uchiyama}, {Takahashi}, \& {HEGRA Collaboration}}]{tevj2032}
{Aharonian}, F., {Akhperjanian}, A., {Beilicke}, M., {et~al.} 2002, \aap, 393,
  L37

\bibitem[{{Aharonian} {et~al.}(2006{\natexlab{a}}){Aharonian}, {Akhperjanian},
  {Bazer-Bachi}, {Beilicke}, {Benbow}, {Berge}, {Bernl{\"o}hr}, {Boisson},
  {Bolz}, {Borrel}, {Braun}, {Breitling}, {Brown}, {B{\"u}hler},
  {B{\"u}sching}, {Carrigan}, {Chadwick}, {Chounet}, {Cornils}, {Costamante},
  {Degrange}, {Dickinson}, {Djannati-Ata{\"i}}, {O'C.~Drury}, {Dubus},
  {Egberts}, {Emmanoulopoulos}, {Espigat}, {Feinstein}, {Ferrero}, {Fiasson},
  {Fontaine}, {Funk}, {Funk}, {Gallant}, {Giebels}, {Glicenstein}, {Goret},
  {Hadjichristidis}, {Hauser}, {Hauser}, {Heinzelmann}, {Henri}, {Hermann},
  {Hinton}, {Hofmann}, {Holleran}, {Horns}, {Jacholkowska}, {de Jager},
  {Kh{\'e}lifi}, {Komin}, {Konopelko}, {Kosack}, {Latham}, {Le Gallou},
  {Lemi{\`e}re}, {Lemoine-Goumard}, {Lohse}, {Martin}, {Martineau-Huynh},
  {Marcowith}, {Masterson}, {McComb}, {de Naurois}, {Nedbal}, {Nolan},
  {Noutsos}, {Orford}, {Osborne}, {Ouchrif}, {Panter}, {Pelletier}, {Pita},
  {P{\"u}hlhofer}, {Punch}, {Raubenheimer}, {Raue}, {Rayner}, {Reimer},
  {Reimer}, {Ripken}, {Rob}, {Rolland}, {Rowell}, {Sahakian}, {Saug{\'e}},
  {Schlenker}, {Schlickeiser}, {Schwanke}, {Sol}, {Spangler}, {Spanier},
  {Steenkamp}, {Stegmann}, {Superina}, {Tavernet}, {Terrier}, {Th{\'e}oret},
  {Tluczykont}, {van Eldik}, {Vasileiadis}, {Venter}, {Vincent}, {V{\"o}lk},
  {Wagner}, \& {Ward}}]{hesscrab}
{Aharonian}, F., {Akhperjanian}, A.~G., {Bazer-Bachi}, A.~R., {et~al.}
  2006{\natexlab{a}}, \aap, 457, 899

\bibitem[{{Aharonian} {et~al.}(2006{\natexlab{b}}){Aharonian}, {Akhperjanian},
  {Bazer-Bachi}, {Beilicke}, {Benbow}, {Berge}, {Bernl{\"o}hr}, {Boisson},
  {Bolz}, {Borrel}, {Braun}, {Breitling}, {Brown}, {Chadwick}, {Chounet},
  {Cornils}, {Costamante}, {Degrange}, {Dickinson}, {Djannati-Ata{\"i}},
  {Drury}, {Dubus}, {Emmanoulopoulos}, {Espigat}, {Feinstein}, {Fontaine},
  {Fuchs}, {Funk}, {Gallant}, {Giebels}, {Gillessen}, {Glicenstein}, {Goret},
  {Hadjichristidis}, {Hauser}, {Heinzelmann}, {Henri}, {Hermann}, {Hinton},
  {Hofmann}, {Holleran}, {Horns}, {Jacholkowska}, {de Jager}, {Kh{\'e}lifi},
  {Komin}, {Konopelko}, {Latham}, {Le Gallou}, {Lemi{\`e}re},
  {Lemoine-Goumard}, {Leroy}, {Lohse}, {Martin}, {Martineau-Huynh},
  {Marcowith}, {Masterson}, {McComb}, {de Naurois}, {Nolan}, {Noutsos},
  {Orford}, {Osborne}, {Ouchrif}, {Panter}, {Pelletier}, {Pita},
  {P{\"u}hlhofer}, {Punch}, {Raubenheimer}, {Raue}, {Raux}, {Rayner}, {Reimer},
  {Reimer}, {Ripken}, {Rob}, {Rolland}, {Rowell}, {Sahakian}, {Saug{\'e}},
  {Schlenker}, {Schlickeiser}, {Schuster}, {Schwanke}, {Siewert}, {Sol},
  {Spangler}, {Steenkamp}, {Stegmann}, {Tavernet}, {Terrier}, {Th{\'e}oret},
  {Tluczykont}, {Vasileiadis}, {Venter}, {Vincent}, {V{\"o}lk}, \&
  {Wagner}}]{hessscanII}
{Aharonian}, F., {Akhperjanian}, A.~G., {Bazer-Bachi}, A.~R., {et~al.}
  2006{\natexlab{b}}, \apj, 636, 777

\bibitem[{{Aharonian} {et~al.}(2006{\natexlab{c}}){Aharonian}, {Akhperjanian},
  {Bazer-Bachi}, {Beilicke}, {Benbow}, {Berge}, {Bernl{\"o}hr}, {Boisson},
  {Bolz}, {Borrel}, {Braun}, {Breitling}, {Brown}, {Chadwick}, {Chounet},
  {Cornils}, {Costamante}, {Degrange}, {Dickinson}, {Djannati-Ata{\"i}},
  {Drury}, {Dubus}, {Emmanoulopoulos}, {Espigat}, {Feinstein}, {Fontaine},
  {Fuchs}, {Funk}, {Gallant}, {Giebels}, {Gillessen}, {Glicenstein}, {Goret},
  {Hadjichristidis}, {Hauser}, {Hauser}, {Heinzelmann}, {Henri}, {Hermann},
  {Hinton}, {Hofmann}, {Holleran}, {Horns}, {Jacholkowska}, {de Jager},
  {Kh{\'e}lifi}, {Klages}, {Komin}, {Konopelko}, {Latham}, {Le Gallou},
  {Lemi{\`e}re}, {Lemoine-Goumard}, {Leroy}, {Lohse}, {Marcowith}, {Martin},
  {Martineau-Huynh}, {Masterson}, {McComb}, {de Naurois}, {Nolan}, {Noutsos},
  {Orford}, {Osborne}, {Ouchrif}, {Panter}, {Pelletier}, {Pita},
  {P{\"u}hlhofer}, {Punch}, {Raubenheimer}, {Raue}, {Raux}, {Rayner}, {Reimer},
  {Reimer}, {Ripken}, {Rob}, {Rolland}, {Rowell}, {Sahakian}, {Saug{\'e}},
  {Schlenker}, {Schlickeiser}, {Schuster}, {Schwanke}, {Siewert}, {Sol},
  {Spangler}, {Steenkamp}, {Stegmann}, {Tavernet}, {Terrier}, {Th{\'e}oret},
  {Tluczykont}, {van Eldik}, {Vasileiadis}, {Venter}, {Vincent}, {V{\"o}lk}, \&
  {Wagner}}]{hessgalcen}
{Aharonian}, F., {Akhperjanian}, A.~G., {Bazer-Bachi}, A.~R., {et~al.}
  2006{\natexlab{c}}, \nat, 439, 695

\bibitem[{{Aharonian} {et~al.}(2007){Aharonian}, {Akhperjanian}, {Bazer-Bachi},
  {Beilicke}, {Benbow}, {Berge}, {Bernl{\"o}hr}, {Boisson}, {Bolz}, {Borrel},
  {Braun}, {Brion}, {Brown}, {B{\"u}hler}, {B{\"u}sching}, {Boutelier},
  {Carrigan}, {Chadwick}, {Chounet}, {Coignet}, {Cornils}, {Costamante},
  {Degrange}, {Dickinson}, {Djannati-Ata{\"i}}, {Drury}, {Dubus}, {Egberts},
  {Emmanoulopoulos}, {Espigat}, {Farnier}, {Feinstein}, {Ferrero}, {Fiasson},
  {Fontaine}, {Funk}, {Funk}, {F{\"u}{\ss}ling}, {Gallant}, {Giebels},
  {Glicenstein}, {Gl{\"u}ck}, {Goret}, {Hadjichristidis}, {Hauser}, {Hauser},
  {Heinzelmann}, {Henri}, {Hermann}, {Hinton}, {Hoffmann}, {Hofmann},
  {Holleran}, {Hoppe}, {Horns}, {Jacholkowska}, {de Jager}, {Kendziorra},
  {Kerschhaggl}, {Kh{\'e}lifi}, {Komin}, {Kosack}, {Lamanna}, {Latham}, {Le
  Gallou}, {Lemi{\`e}re}, {Lemoine-Goumard}, {Lohse}, {Martin},
  {Martineau-Huynh}, {Marcowith}, {Masterson}, {Maurin}, {McComb}, {Moulin},
  {de Naurois}, {Nedbal}, {Nolan}, {Noutsos}, {Olive}, {Orford}, {Osborne},
  {Panter}, {Pelletier}, {Petrucci}, {Pita}, {P{\"u}hlhofer}, {Punch},
  {Ranchon}, {Raubenheimer}, {Raue}, {Rayner}, {Reimer}, {Reimer}, {Ripken},
  {Rob}, {Rolland}, {Rosier-Lees}, {Rowell}, {Sahakian}, {Santangelo},
  {Saug{\'e}}, {Schlenker}, {Schlickeiser}, {Schr{\"o}der}, {Schwanke},
  {Schwarzburg}, {Schwemmer}, {Shalchi}, {Sol}, {Spangler}, {Spanier},
  {Steenkamp}, {Stegmann}, {Superina}, {Tam}, {Tavernet}, {Terrier},
  {Tluczykont}, {van Eldik}, {Vasileiadis}, {Venter}, {Vialle}, {Vincent},
  {V{\"o}lk}, {Wagner}, \& {Ward}}]{hesswd2}
{Aharonian}, F., {Akhperjanian}, A.~G., {Bazer-Bachi}, A.~R., {et~al.} 2007,
  \aap, 467, 1075

\bibitem[{{Aharonian} \& {Atoyan}(1981)}]{aharonian:1981}
{Aharonian}, F.~A. \& {Atoyan}, A.~M. 1981, \apss, 79, 321

\bibitem[{{Aleksi{\'c}} {et~al.}(2013){Aleksi{\'c}}, {Antonelli}, {Antoranz},
  {Asensio}, {Barres de Almeida}, {Barrio}, {Becerra Gonz{\'a}lez}, {Bednarek},
  {Berger}, {Bernardini}, {Biland}, {Blanch}, {Bock}, {Boller}, {Bonnoli},
  {Borla Tridon}, {Bretz}, {Carmona}, {Carosi}, {Colin}, {Colombo},
  {Contreras}, {Cortina}, {Cossio}, {Covino}, {Da Vela}, {Dazzi}, {De Angelis},
  {De Caneva}, {De Cea del Pozo}, {De Lotto}, {Delgado Mendez}, {Diago Ortega},
  {Doert}, {Dominis Prester}, {Dorner}, {Doro}, {Eisenacher}, {Elsaesser},
  {Ferenc}, {Fonseca}, {Font}, {Fruck}, {Garc{\'{\i}}a L{\'o}pez},
  {Garczarczyk}, {Garrido Terrats}, {Gaug}, {Giavitto}, {Godinovi{\'c}},
  {Gonz{\'a}lez Mu{\~n}oz}, {Gozzini}, {Hadamek}, {Hadasch}, {H{\"a}fner},
  {Herrero}, {Hose}, {Hrupec}, {Huber}, {Jankowski}, {Jogler}, {Kadenius},
  {Klepser}, {Knoetig}, {Kr{\"a}henb{\"u}hl}, {Krause}, {Kushida}, {La
  Barbera}, {Lelas}, {Leonardo}, {Lewandowska}, {Lindfors}, {Lombardi},
  {L{\'o}pez}, {L{\'o}pez-Coto}, {L{\'o}pez-Oramas}, {Lorenz}, {Makariev},
  {Maneva}, {Mankuzhiyil}, {Mannheim}, {Maraschi}, {Marcote}, {Mariotti},
  {Mart{\'{\i}}nez}, {Mazin}, {Meucci}, {Miranda}, {Mirzoyan}, {Mold{\'o}n},
  {Moralejo}, {Munar-Adrover}, {Niedzwiecki}, {Nieto}, {Nilsson}, {Nowak},
  {Orito}, {Paiano}, {Palatiello}, {Paneque}, {Paoletti}, {Paredes}, {Partini},
  {Persic}, {Pilia}, {Pochon}, {Prada}, {Prada Moroni}, {Prandini}, {Puljak},
  {Reichardt}, {Reinthal}, {Rhode}, {Rib{\'o}}, {Rico}, {R{\"u}gamer},
  {Saggion}, {Saito}, {Saito}, {Salvati}, {Satalecka}, {Scalzotto}, {Scapin},
  {Schultz}, {Schweizer}, {Shore}, {Sillanp{\"a}{\"a}}, {Sitarek}, {Snidaric},
  {Sobczynska}, {Spanier}, {Spiro}, {Stamatescu}, {Stamerra}, {Steinke},
  {Storz}, {Sun}, {Suri{\'c}}, {Takalo}, {Takami}, {Tavecchio}, {Temnikov},
  {Terzi{\'c}}, {Tescaro}, {Teshima}, {Tibolla}, {Torres}, {Toyama}, {Treves},
  {Uellenbeck}, {Vogler}, {Wagner}, {Weitzel}, {Zabalza}, {Zandanel}, {Zanin},
  {Rea}, \& {Backes}}]{magic-magnetars}
{Aleksi{\'c}}, J., {Antonelli}, L.~A., {Antoranz}, P., {et~al.} 2013, \aap,
  549, A23

\bibitem[{{Arons}(2003)}]{arons:2003}
{Arons}, J. 2003, \apj, 589, 871

\bibitem[{{Becherini} {et~al.}(2011){Becherini}, {Djannati-Ata{\"i}},
  {Marandon}, {Punch}, \& {Pita}}]{hessMVA}
{Becherini}, Y., {Djannati-Ata{\"i}}, A., {Marandon}, V., {Punch}, M., \&
  {Pita}, S. 2011, Astroparticle Physics, 34, 858

\bibitem[{{Bednarek}(2007)}]{bednarek:2007}
{Bednarek}, W. 2007, \mnras, 382, 367

\bibitem[{{Berge} {et~al.}(2007){Berge}, {Funk}, \& {Hinton}}]{berge:2007}
{Berge}, D., {Funk}, S., \& {Hinton}, J. 2007, \aap, 466, 1219

\bibitem[{{Bibby} {et~al.}(2008){Bibby}, {Crowther}, {Furness}, \&
  {Clark}}]{bibby:2008}
{Bibby}, J.~L., {Crowther}, P.~A., {Furness}, J.~P., \& {Clark}, J.~S. 2008,
  \mnras, 386, L23

\bibitem[{{Clark} {et~al.}(2005){Clark}, {Larionov}, \&
  {Arkharov}}]{clark:2005}
{Clark}, J.~S., {Larionov}, V.~M., \& {Arkharov}, A. 2005, \aap, 435, 239

\bibitem[{{Clark} {et~al.}(2008){Clark}, {Muno}, {Negueruela}, {Dougherty},
  {Crowther}, {Goodwin}, \& {de Grijs}}]{clark:2008}
{Clark}, J.~S., {Muno}, M.~P., {Negueruela}, I., {et~al.} 2008, \aap, 477, 147

\bibitem[{{Corbel} \& {Eikenberry}(2004)}]{corbel:2004}
{Corbel}, S. \& {Eikenberry}, S.~S. 2004, \aap, 419, 191

\bibitem[{{Crutcher} {et~al.}(2010){Crutcher}, {Wandelt}, {Heiles},
  {Falgarone}, \& {Troland}}]{crutcher:2010}
{Crutcher}, R.~M., {Wandelt}, B., {Heiles}, C., {Falgarone}, E., \& {Troland},
  T.~H. 2010, \apj, 725, 466

\bibitem[{{de Naurois} \& {Rolland}(2009)}]{hessmodel}
{de Naurois}, M. \& {Rolland}, L. 2009, Astroparticle Physics, 32, 231

\bibitem[{{Domingo-Santamar{\'{\i}}a} \& {Torres}(2006)}]{domingo:2006}
{Domingo-Santamar{\'{\i}}a}, E. \& {Torres}, D.~F. 2006, \aap, 448, 613

\bibitem[{{Duncan} \& {Thompson}(1992)}]{duncan:1992}
{Duncan}, R.~C. \& {Thompson}, C. 1992, \apjl, 392, L9

\bibitem[{{Edwards} {et~al.}(2011){Edwards}, {Bandyopadhyay}, {Eikenberry},
  {Mikles}, \& {Moon}}]{edwards:2011}
{Edwards}, M.~L., {Bandyopadhyay}, R.~M., {Eikenberry}, S.~S., {Mikles}, V.~J.,
  \& {Moon}, D.-S. 2011, in IAU Symposium, Vol. 272, IAU Symposium, ed.
  C.~{Neiner}, G.~{Wade}, G.~{Meynet}, \& G.~{Peters}, 606--607

\bibitem[{{Eichler} \& {Usov}(1993)}]{eichler:1993}
{Eichler}, D. \& {Usov}, V. 1993, \apj, 402, 271

\bibitem[{{Eikenberry} {et~al.}(2001){Eikenberry}, {Garske}, {Hu}, {Jackson},
  {Patel}, {Barry}, {Colonno}, \& {Houck}}]{eikenberry:2001}
{Eikenberry}, S.~S., {Garske}, M.~A., {Hu}, D., {et~al.} 2001, \apjl, 563, L133

\bibitem[{{Eikenberry} {et~al.}(2004){Eikenberry}, {Matthews}, {LaVine},
  {Garske}, {Hu}, {Jackson}, {Patel}, {Barry}, {Colonno}, {Houck}, {Wilson},
  {Corbel}, \& {Smith}}]{eikenberry:2004}
{Eikenberry}, S.~S., {Matthews}, K., {LaVine}, J.~L., {et~al.} 2004, \apj, 616,
  506

\bibitem[{{Esposito} {et~al.}(2007){Esposito}, {Mereghetti}, {Tiengo}, {Zane},
  {Turolla}, {G{\"o}tz}, {Rea}, {Kawai}, {Ueno}, {Israel}, {Stella}, \&
  {Feroci}}]{esposito:2007}
{Esposito}, P., {Mereghetti}, S., {Tiengo}, A., {et~al.} 2007, \aap, 476, 321

\bibitem[{{Figer} {et~al.}(2005){Figer}, {Najarro}, {Geballe}, {Blum}, \&
  {Kudritzki}}]{figer:2005}
{Figer}, D.~F., {Najarro}, F., {Geballe}, T.~R., {Blum}, R.~D., \& {Kudritzki},
  R.~P. 2005, \apjl, 622, L49

\bibitem[{{Figer} {et~al.}(2004){Figer}, {Najarro}, \&
  {Kudritzki}}]{figer:2004}
{Figer}, D.~F., {Najarro}, F., \& {Kudritzki}, R.~P. 2004, \apjl, 610, L109

\bibitem[{{Fuchs} {et~al.}(1999){Fuchs}, {Mirabel}, {Chaty}, {Claret},
  {Cesarsky}, \& {Cesarsky}}]{fuchs:1999}
{Fuchs}, Y., {Mirabel}, F., {Chaty}, S., {et~al.} 1999, \aap, 350, 891

\bibitem[{{Gabici} {et~al.}(2007){Gabici}, {Aharonian}, \&
  {Blasi}}]{gabici:2007}
{Gabici}, S., {Aharonian}, F.~A., \& {Blasi}, P. 2007, \apss, 309, 365

\bibitem[{{Gaensler} {et~al.}(2001){Gaensler}, {Slane}, {Gotthelf}, \&
  {Vasisht}}]{gaensler:2001}
{Gaensler}, B.~M., {Slane}, P.~O., {Gotthelf}, E.~V., \& {Vasisht}, G. 2001,
  \apj, 559, 963

\bibitem[{{G{\"o}tz} {et~al.}(2007){G{\"o}tz}, {Mereghetti}, \&
  {Hurley}}]{goetz:2007}
{G{\"o}tz}, D., {Mereghetti}, S., \& {Hurley}, K. 2007, \apss, 308, 51

\bibitem[{{Halpern} \& {Gotthelf}(2010)}]{halpern:2010}
{Halpern}, J.~P. \& {Gotthelf}, E.~V. 2010, \apj, 725, 1384

\bibitem[{{Harding} \& {Lai}(2006)}]{harding:2006}
{Harding}, A.~K. \& {Lai}, D. 2006, Reports on Progress in Physics, 69, 2631

\bibitem[{{Hurley} {et~al.}(2005){Hurley}, {Boggs}, {Smith}, {Duncan}, {Lin},
  {Zoglauer}, {Krucker}, {Hurford}, {Hudson}, {Wigger}, {Hajdas}, {Thompson},
  {Mitrofanov}, {Sanin}, {Boynton}, {Fellows}, {von Kienlin}, {Lichti}, {Rau},
  \& {Cline}}]{hurley:2005}
{Hurley}, K., {Boggs}, S.~E., {Smith}, D.~M., {et~al.} 2005, \nat, 434, 1098

\bibitem[{{Ioka} {et~al.}(2005){Ioka}, {Razzaque}, {Kobayashi}, \&
  {M{\'e}sz{\'a}ros}}]{ioka:2005}
{Ioka}, K., {Razzaque}, S., {Kobayashi}, S., \& {M{\'e}sz{\'a}ros}, P. 2005,
  \apj, 633, 1013

\bibitem[{{Kaplan} {et~al.}(2002){Kaplan}, {Fox}, {Kulkarni}, {Gotthelf},
  {Vasisht}, \& {Frail}}]{kaplan:2002}
{Kaplan}, D.~L., {Fox}, D.~W., {Kulkarni}, S.~R., {et~al.} 2002, \apj, 564, 935

\bibitem[{{Kargaltsev} {et~al.}(2013){Kargaltsev}, {Rangelov}, \&
  {Pavlov}}]{kargaltsev:2013}
{Kargaltsev}, O., {Rangelov}, B., \& {Pavlov}, G. 2013, {Pulsar-Wind Nebulae as
  a Dominant Population of Galactic VHE Sources}, 359--406

\bibitem[{{Kniazev} {et~al.}(2015){Kniazev}, {Gvaramadze}, \&
  {Berdnikov}}]{kniazev:2015}
{Kniazev}, A.~Y., {Gvaramadze}, V.~V., \& {Berdnikov}, L.~N. 2015, \mnras, 449,
  L60

\bibitem[{{Kulkarni} {et~al.}(1994){Kulkarni}, {Frail}, {Kassim}, {Murakami},
  \& {Vasisht}}]{kulkarni:1994}
{Kulkarni}, S.~R., {Frail}, D.~A., {Kassim}, N.~E., {Murakami}, T., \&
  {Vasisht}, G. 1994, \nat, 368, 129

\bibitem[{{Kulkarni} {et~al.}(1995){Kulkarni}, {Matthews}, {Neugebauer},
  {Reid}, {van Kerkwijk}, \& {Vasisht}}]{kulkarni:1995}
{Kulkarni}, S.~R., {Matthews}, K., {Neugebauer}, G., {et~al.} 1995, \apjl, 440,
  L61

\bibitem[{{Laros} {et~al.}(1986){Laros}, {Fenimore}, {Fikani}, {Klebesadel}, \&
  {Barat}}]{laros:1986}
{Laros}, J.~G., {Fenimore}, E.~E., {Fikani}, M.~M., {Klebesadel}, R.~W., \&
  {Barat}, C. 1986, \nat, 322, 152

\bibitem[{{Li} \& {Ma}(1983)}]{Li:1983}
{Li}, T.-P. \& {Ma}, Y.-Q. 1983, \apj, 272, 317

\bibitem[{{Liu} {et~al.}(2010){Liu}, {Wu}, \& {Lu}}]{liu:2010}
{Liu}, X.-W., {Wu}, X.-F., \& {Lu}, T. 2010, \na, 15, 292

\bibitem[{{Mattana} {et~al.}(2009){Mattana}, {Falanga}, {G{\"o}tz}, {Terrier},
  {Esposito}, {Pellizzoni}, {De Luca}, {Marandon}, {Goldwurm}, \&
  {Caraveo}}]{mattana:2009}
{Mattana}, F., {Falanga}, M., {G{\"o}tz}, D., {et~al.} 2009, \apj, 694, 12

\bibitem[{{McClure-Griffiths} \& {Gaensler}(2005)}]{mcclure:2005}
{McClure-Griffiths}, N.~M. \& {Gaensler}, B.~M. 2005, \apjl, 630, L161

\bibitem[{{Mereghetti}(2011)}]{mereghetti:2011}
{Mereghetti}, S. 2011, Advances in Space Research, 47, 1317

\bibitem[{{Mereghetti} {et~al.}(2007){Mereghetti}, {Esposito}, \&
  {Tiengo}}]{mereghetti:2007}
{Mereghetti}, S., {Esposito}, P., \& {Tiengo}, A. 2007, \apss, 308, 13

\bibitem[{{Montmerle}(1979)}]{montmerle:1979}
{Montmerle}, T. 1979, \apj, 231, 95

\bibitem[{{Muno} {et~al.}(2007){Muno}, {Gaensler}, {Clark}, {de Grijs},
  {Pooley}, {Stevens}, \& {Portegies Zwart}}]{muno:2007}
{Muno}, M.~P., {Gaensler}, B.~M., {Clark}, J.~S., {et~al.} 2007, \mnras, 378,
  L44

\bibitem[{{Nakagawa} {et~al.}(2009{\natexlab{a}}){Nakagawa}, {Mihara},
  {Yoshida}, {Yamaoka}, {Sugita}, {Murakami}, {Yonetoku}, {Suzuki}, {Nakajima},
  {Tashiro}, \& {Nakazawa}}]{nakagawa:2009b}
{Nakagawa}, Y.~E., {Mihara}, T., {Yoshida}, A., {et~al.} 2009{\natexlab{a}},
  \pasj, 61, 387

\bibitem[{{Nakagawa} {et~al.}(2009{\natexlab{b}}){Nakagawa}, {Yoshida},
  {Yamaoka}, \& {Shibazaki}}]{nakagawa:2009a}
{Nakagawa}, Y.~E., {Yoshida}, A., {Yamaoka}, K., \& {Shibazaki}, N.
  2009{\natexlab{b}}, \pasj, 61, 109

\bibitem[{{Naz{\'e}} {et~al.}(2012){Naz{\'e}}, {Rauw}, \&
  {Hutsem{\'e}kers}}]{naze:2012}
{Naz{\'e}}, Y., {Rauw}, G., \& {Hutsem{\'e}kers}, D. 2012, \aap, 538, A47

\bibitem[{{Olausen} \& {Kaspi}(2014)}]{olausen:2014}
{Olausen}, S.~A. \& {Kaspi}, V.~M. 2014, \apjs, 212, 6

\bibitem[{{Ouyed} {et~al.}(2007){Ouyed}, {Leahy}, \& {Niebergal}}]{ouyed:2007}
{Ouyed}, R., {Leahy}, D., \& {Niebergal}, B. 2007, \aap, 473, 357

\bibitem[{{Paczynski}(1992)}]{paczynski:1992}
{Paczynski}, B. 1992, \actaa, 42, 145

\bibitem[{{Protheroe} {et~al.}(2008){Protheroe}, {Ott}, {Ekers}, {Jones}, \&
  {Crocker}}]{protheroe:2008}
{Protheroe}, R.~J., {Ott}, J., {Ekers}, R.~D., {Jones}, D.~I., \& {Crocker},
  R.~M. 2008, \mnras, 390, 683

\bibitem[{{Rahoui} {et~al.}(2009){Rahoui}, {Chaty}, \& {Lagage}}]{rahoui:2009}
{Rahoui}, F., {Chaty}, S., \& {Lagage}, P.-O. 2009, \aap, 493, 119

\bibitem[{{Reimer} {et~al.}(2006){Reimer}, {Pohl}, \& {Reimer}}]{reimer:2006}
{Reimer}, A., {Pohl}, M., \& {Reimer}, O. 2006, \apj, 644, 1118

\bibitem[{{Reitberger} {et~al.}(2015){Reitberger}, {Reimer}, {Reimer}, \&
  {Takahashi}}]{reitberger:2015}
{Reitberger}, K., {Reimer}, A., {Reimer}, O., \& {Takahashi}, H. 2015, \aap,
  577, A100

\bibitem[{{Rowell}(2003)}]{template}
{Rowell}, G.~P. 2003, \aap, 410, 389

\bibitem[{{Svirski} {et~al.}(2011){Svirski}, {Nakar}, \& {Ofek}}]{svirski:2011}
{Svirski}, G., {Nakar}, E., \& {Ofek}, E.~O. 2011, \mnras, 415, 2485

\bibitem[{{Tavani} {et~al.}(2009){Tavani}, {Sabatini}, {Pian}, {Bulgarelli},
  {Caraveo}, {Viotti}, {Corcoran}, {Giuliani}, {Pittori}, {Verrecchia},
  {Vercellone}, {Mereghetti}, {Argan}, {Barbiellini}, {Boffelli}, {Cattaneo},
  {Chen}, {Cocco}, {D'Ammando}, {Costa}, {DeParis}, {Del Monte}, {Di Cocco},
  {Donnarumma}, {Evangelista}, {Ferrari}, {Feroci}, {Fiorini}, {Froysland},
  {Fuschino}, {Galli}, {Gianotti}, {Labanti}, {Lapshov}, {Lazzarotto},
  {Lipari}, {Longo}, {Marisaldi}, {Mastropietro}, {Morelli}, {Moretti},
  {Morselli}, {Pacciani}, {Pellizzoni}, {Perotti}, {Piano}, {Picozza}, {Pilia},
  {Porrovecchio}, {Pucella}, {Prest}, {Rapisarda}, {Rappoldi}, {Rubini},
  {Soffitta}, {Trifoglio}, {Trois}, {Vallazza}, {Vittorini}, {Zambra},
  {Zanello}, {Santolamazza}, {Giommi}, {Colafrancesco}, {Antonelli}, \&
  {Salotti}}]{etacaragile}
{Tavani}, M., {Sabatini}, S., {Pian}, E., {et~al.} 2009, \apjl, 698, L142

\bibitem[{{Tendulkar} {et~al.}(2012){Tendulkar}, {Cameron}, \&
  {Kulkarni}}]{tendulkar:2012}
{Tendulkar}, S.~P., {Cameron}, P.~B., \& {Kulkarni}, S.~R. 2012, \apj, 761, 76

\bibitem[{{Tibolla} {et~al.}(2009){Tibolla}, {Komin}, {Kosack}, \&
  {Naumann-Godo}}]{tibolla:2009}
{Tibolla}, O., {Komin}, N., {Kosack}, K., \& {Naumann-Godo}, M. 2009, in
  American Institute of Physics Conference Series, Vol. 1112, , 233--237

\bibitem[{{van Kerkwijk} {et~al.}(1995){van Kerkwijk}, {Kulkarni}, {Matthews},
  \& {Neugebauer}}]{vankerwijk:1995}
{van Kerkwijk}, M.~H., {Kulkarni}, S.~R., {Matthews}, K., \& {Neugebauer}, G.
  1995, \apjl, 444, L33

\bibitem[{{Vigan{\`o}} {et~al.}(2014){Vigan{\`o}}, {Rea}, {Esposito},
  {Mereghetti}, {Israel}, {Tiengo}, {Turolla}, {Zane}, \&
  {Stella}}]{vigano:2014}
{Vigan{\`o}}, D., {Rea}, N., {Esposito}, P., {et~al.} 2014, Journal of High
  Energy Astrophysics, 3, 41

\bibitem[{{Voelk} \& {Forman}(1982)}]{volk:1982}
{Voelk}, H.~J. \& {Forman}, M. 1982, \apj, 253, 188

\bibitem[{{Woods} {et~al.}(2007){Woods}, {Kouveliotou}, {Finger}, {G{\"o}{\v
  g}{\"u}{\c s}}, {Wilson}, {Patel}, {Hurley}, \& {Swank}}]{woods:2007}
{Woods}, P.~M., {Kouveliotou}, C., {Finger}, M.~H., {et~al.} 2007, \apj, 654,
  470

\bibitem[{{Younes} {et~al.}(2015){Younes}, {Kouveliotou}, \&
  {Kaspi}}]{younes:2015}
{Younes}, G., {Kouveliotou}, C., \& {Kaspi}, V.~M. 2015, \apj, 809, 165

\bibitem[{Zhang(2003)}]{zhang:2003}
Zhang, B. 2003, in {Proceedings, International Workshop on Strong Magnetic
  Fields and Neutron Star}, 83

\bibitem[{{Zhang} {et~al.}(2003){Zhang}, {Dai}, {M{\'e}sz{\'a}ros}, {Waxman},
  \& {Harding}}]{zhang:2003b}
{Zhang}, B., {Dai}, Z.~G., {M{\'e}sz{\'a}ros}, P., {Waxman}, E., \& {Harding},
  A.~K. 2003, \apj, 595, 346

\end{thebibliography}
\bibliographystyle{aa}

\clearpage

\appendix

\section{Additional online figures \& tables} 

Additional online figures and tables are given here.

\begin{figure}[th]
  \centering
  \includegraphics[width=0.5\textwidth]{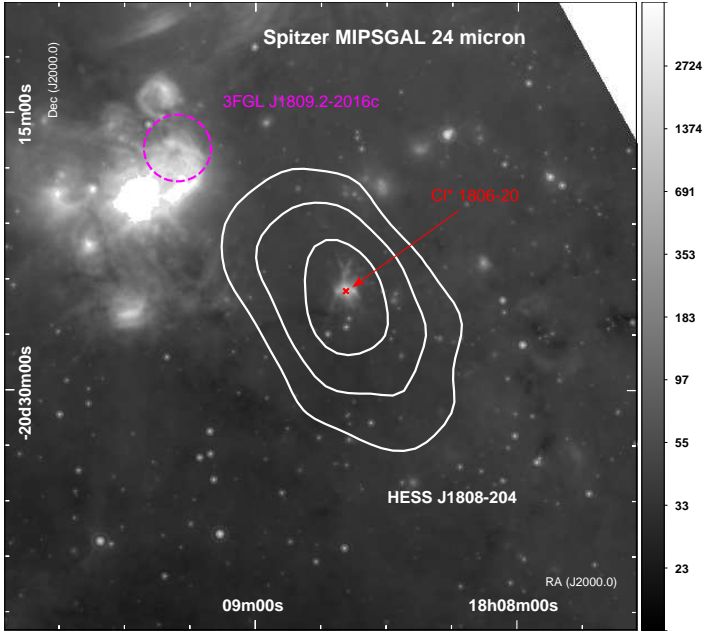}
  \caption{{\em Spitzer} MIPSGAL 24 $\mu$m image in MJy/sr units with HESS\,J1808$-$204 excess significance contours 
    (6, 5, 4 $\sigma$ as for Fig.~\ref{fig:tevimage}) overlaid as solid white lines. 
    Locations of the stellar cluster Cl*\,1806$-$20 (containing
    SGR1806$-$20 and LBV\,1806$-$20) and the {\em Fermi-LAT} GeV source 3FGL\,J1809.2$-$2016c (68\% location error) are indicated. 
    The bright infrared feature to the north-east towards 
    the {\em Fermi-LAT} source is the W31 giant HII star formation complex.}
  \label{fig:ir_image}
\end{figure}
\begin{table}[b] 
  \centering
  \begin{tabular}{ccccc}\\ \hline
     Events              &          & $^1\alpha$ & $^2S$ & Excess events \\
                         &          &            &       & ($N-\alpha N_b$) \\ \hline
    \multicolumn{5}{c}{--- Faint Cuts ---}\\
     On ($N$)            &   3128 \\ 
     Off ($N_b$)         &  16758  &  0.162   & 7.1$\sigma$  & 412.8 \\
     \multicolumn{5}{c}{--- Standard Cuts ---}\\
     On ($N$)            &   5147 \\  
     Off ($N_b$)         &  26470  &  0.174   & 7.4$\sigma$  & 553.4 \\ \hline
     \multicolumn{5}{l}{\scriptsize 1. Normalisation between on source and background regions}\\
     \multicolumn{5}{l}{\scriptsize 2. Statistical significance from Eq. 17 of \citet{Li:1983}.}\\
   \end{tabular}
   \caption{Event statistics of HESS\,J1808$-$204 for events within a radius 0.2$^\circ$ of the fitted position 
     using the reflected background model for both {\em faint} and {\em standard} cuts analyses.}
   \label{tab:statistics}
 \end{table}

 \begin{table}[b]
   \centering
   \begin{tabular}{cccc}\\ \hline
     E (TeV)  & $F$                    & $F$ (down)  &  $F$ (up)\\ 
              & \multicolumn{3}{c}{(photons\,cm$^{-2}$\,s$^{-1}$\,TeV$^{-1}$)} \\ \hline
     0.34 & 3.38$\times 10^{-12}$ & 2.39$\times 10^{-12}$ & 4.39$\times 10^{-12}$ \\
     0.78 & 5.17$\times 10^{-13}$ & 3.95$\times 10^{-13}$ & 6.42$\times 10^{-13}$ \\
     2.04 & 5.10$\times 10^{-14}$ & 3.26$\times 10^{-14}$ & 7.00$\times 10^{-14}$ \\
     5.35 & 6.36$\times 10^{-15}$ & 2.80$\times 10^{-15}$ & 1.02$\times 10^{-14}$ \\ \hline  
   \end{tabular}
   \caption{VHE spectral fluxes $F$ and 68\% statistical error limits of HESS\,J1808$-$204 for events within a 
     radius 0.2$^\circ$ of the fitted position 
     using the reflected background model and {\em faint} cuts analyses.}
   \label{tab:spectrum}
 \end{table}

\end{document}